\documentclass[twocolumn,astrosymb,usenatbib]{aastex7}

\usepackage{color, soul}
\usepackage{verbatim}

\newcommand{\fe}[1] {\textcolor{cyan}{#1 \textsc{/FE/}}}

\newcommand{\al}[1] {\textcolor{blue}{#1 \textsc{/AL/}}}

\begin{document}

\title[Hypervelocity stars from globular clusters]{The ejection and detectability of high- and hyper-velocity stars by compact object binaries in globular clusters}






\author[0000-0002-4521-4809]{Fraser A. Evans}
\affiliation{David A. Dunlap Department of Astronomy and Astrophysics, University of Toronto, 50 St. George Street, Toronto, ON, M5S 3H4, Canada}
\affiliation{Dunlap Institute for Astronomy \& Astrophysics, University of Toronto, 50 St. George Street, Toronto, ON, M5S 3H4, Canada}
\email[show]{fraser.evans@utoronto.ca}  

\author[0000-0002-0444-8502]{Steffani M. Grondin} 
\affiliation{David A. Dunlap Department of Astronomy and Astrophysics, University of Toronto, 50 St. George Street, Toronto, ON, M5S 3H4, Canada}
\email{steffani.grondin@utoronto.ca}  

\author[0000-0001-9582-881X]{Claire S. Ye}
\affiliation{Canadian Institute for Theoretical Astrophysics, University of Toronto, 60 St. George Street, Toronto, Ontario M5S 3H8, Canada}
\email{claireshiye@cita.utoronto.ca}  

\author[0000-0003-3613-0854
]{Jeremy Webb}
\affiliation{Department of Science, Technology and Society, Division of Natural Science, York University, 4700 Keele St, Toronto ON M3J 1P3, Canada}
\email{webbjj@yorku.ca}  

\author[0000-0002-5522-0217]{Alexander Laroche}
\affiliation{David A. Dunlap Department of Astronomy and Astrophysics, University of Toronto, 50 St. George Street, Toronto, ON, M5S 3H4, Canada}
\affiliation{Dunlap Institute for Astronomy \& Astrophysics, University of Toronto, 50 St. George Street, Toronto, ON, M5S 3H4, Canada}
\email{alex.laroche@mail.utoronto.ca}  

\author[0000-0001-6855-442X]{Jo Bovy}
\affiliation{David A. Dunlap Department of Astronomy and Astrophysics, University of Toronto, 50 St. George Street, Toronto, ON, M5S 3H4, Canada}
\affiliation{Dunlap Institute for Astronomy \& Astrophysics, University of Toronto, 50 St. George Street, Toronto, ON, M5S 3H4, Canada}
\email{jo.bovy@utoronto.ca}

\begin{abstract}
The dense cores of Milky Way globular clusters (GCs) play host to a variety of dynamical encounters between stellar objects, which can accelerate stars to velocities high enough to escape the GC. The most extreme examples of these encounters are interactions between single GC stars and binaries including at least one compact object. These interactions can result in ejection velocities of up to several hundred $\mathrm{km \ s^{-1}}$, approaching or even exceeding the escape velocity of the Galaxy itself. In order to study whether these interactions contribute to the Galactic population of \textit{hypervelocity stars} (stars moving faster than the Galactic escape speed), we combine Monte Carlo $N$-body GC simulations, observations of Galactic GCs, and a particle spray code to generate realistic populations of stars which have escaped from Milky Way GCs following star + compact object binary (S+COB) interactions. We find that over the last 500 Myr, S+COB interactions have likely ejected $\sim$6300 stars from Galactic GCs, of which $839_{-67}^{+70}$ have present-day velocities exceeding $500 \; \mathrm{km \ s^{-1}}$. Using mock photometric observations, we find that $290_{-23}^{+28}$ ejected stars are detectable in \textit{Gaia} Data Release 3, however, only $1_{-1}^{+2}$ stars faster than $500 \; \mathrm{km \ s^{-1}}$ are detectable. Even so, we show that observational prospects in the upcoming Legacy Survey of Space and Time are more optimistic, and future detected fast extratidal GC stars will serve as a useful probe of GC cores.
\end{abstract}




\section{Introduction}
Globular clusters (GCs) are spheroidal collections of $\sim 10^{4}-10^{6}$ stars. Primarily located in the Galactic halo, GCs inform large-scale Galaxy evolution as they formed and evolved concurrently with the Milky Way. For instance, GCs provide a variety of insights from past star formation histories \citep{Renaud2017, Forbes2018} to the present-day dark matter content of the Milky Way halo \citep{Varghese2011, Posti2019}. 

GCs in the Galaxy frequently lose mass to the Galactic halo through the process of tidal stripping, wherein the Milky Way tidal field gradually strips the low-mass stars which have been driven to the cluster's outskirts via mass segregation. As a result, stars escape the cluster at speeds just above the cluster escape velocity trail the cluster along its orbit through the Galaxy in extended tidal tails or streams \citep{Combes1999, Johnston1999}. While tidal stripping is effective at removing low-mass stars from a GC, mass-segregation results in a concentration of high-mass objects in the cluster's core \citep[e.g.,][]{Fregeau2002, Heggie2003}. Consequently, dense GC cores are home to a variety of three- and four-body dynamical encounters that occur quite frequently and are capable of ejecting single stars at significant velocities \citep{Poveda1967, Hut1983, Leonard1991}, often fast enough to escape the GC \citep{Weatherford2023, Grondin2023, Grondin2024GEMS}. Of particular interest is the existence of binaries within GC cores in which one or both members are compact objects \citep[white dwarf, neutron star or black hole;][]{Morscher2013, Kremer2018, Giesers2018, Giesers2019,Ye2019}. Interactions between single stars and these \textit{compact object binaries}  (COBs) can result in ejection velocities of several hundred $\mathrm{km \ s^{-1}}$ \citep{Cabrera2023}. The fastest of these escape their host GC with a Galactocentric total velocity above the Milky Way escape velocity \citep[$\sim$600 $\mathrm{km \ s^{-1}}$ for the innermost GCs, falling to $\sim$400 $\mathrm{km \ s^{-1}}$ for the outermost, see][]{Williams2017}. 


The existence of such hypervelocity stars (HVSs), with velocities in excess of the Galactic escape velocity, was first suggested by \citet{Hills1988}, who proposed that dynamical encounters between stellar binaries and Sgr A*, the $4\times10^6 \; \mathrm{M_\odot}$ massive black hole located in the Galactic Centre \citep{Ghez2008, Akiyama2022}, could tidally separate the binary and eject one member at up to several thousand $\mathrm{km \ s^{-1}}$. This so-called \textit{Hills mechanism} \citep[see also e.g.][]{Yu2003, Madigan2009, Sari2010, Rossi2014, Generozov2021} remains to this day the most promising pathway for accelerating stars to extreme velocities.

Since the first discovery of an HVS candidate by \citet{Brown2005}, observational efforts have uncovered several dozen additional HVS candidates \citep{Brown2006, Brown2007, Brown2012, Brown2014, Palladino2014, Zhong2014, ElBadry2023, Burgrasser2024}. The ongoing pursuit of the European Space Agency's \textit{Gaia} mission \citep{Gaia2016} to obtain full 6D positions and velocities for $\sim$tens of millions of Galactic sources and five-parameter astrometry for two billion stars has sparked renewed interest in the fastest Milky Way stars. While \textit{Gaia} data alone have not proven particularly fruitful in uncovering new HVS candidates \citep{Marchetti2018, Hattori2018, Du2019, Marchetti2019, Marchetti2021, Marchetti2022, Liao2023, Verberne2024}, it has been instrumental in re-analysing existing candidates \citep{Brown2018, Irrgang2018, Boubert2018} and interesting new candidates have been uncovered via cross-matching \textit{Gaia} with complementary spectroscopic surveys \citep{Luna2019, Li2021, Li2023, Luna2024, Sun2025} or via spectroscopic follow-up observations of sources identified in \textit{Gaia} \citep[e.g.][]{Bonifacio2024, Caffau2024} .

Due to observational uncertainties, the exact ejection location of most HVS candidates remains unconstrained, with the notable exception of S5-HVS1 \citep{Koposov2020} which can be uncontroversially associated with an origin in the Galactic Centre. The trajectories of many promising candidates, however, are inconsistent with an origin in the Galactic Centre \citep[see e.g.][]{Irrgang2018, Kreuzer2020, Irrgang2021}. Many seem to have been ejected from the Galactic disc \citep{Tillich2009, Huang2017, Li2018, Hattori2019, Irrgang2019} or from outside the inner Galaxy \citep{Marchetti2018, Erkal2019, Huang2021, Prudil2022, Caffau2024}. Unable to harness the gravity of Sgr A*, alternative ejection scenarios must be invoked to explain the high velocities of these objects. Three-body interactions in GCs involving black hole binaries are one such promising alternative -- other potential mechanisms include the ejection of a companion following a core collapse event in a tight binary \citep{Blaauw1961, Tauris1998, Renzo2019, Evans2020}, the ejection of a stripped companion or surviving remnant of a Type Ia supernova \citep{Geier2013, Shen2018, Neuntefel2020, Igoshev2022}, dynamical ejections from young stellar clusters \citep{Oh2016}, the stripping of material from infalling dwarf galaxies \citep{Abadi2009, Piffl2011}, Hills mechanism ejections from massive black holes located in nearby galaxies or dwarf galaxies \citep{Sherwin2008, Boubert2016, Huang2017, Evans2021, Gulzow2024}, and Hills-like ejections from Milky Way star clusters hosting intermediate mass black holes \citep{Gvaramadze2008, Fragione2019}. The population size, stellar properties and kinematics of HVSs can provide valuable insight into a variety of rare and/or difficult-to-observe environments and processes in the Galaxy, including binary evolution and supernova physics \citep{Geier2015, Bauer2019, Portegies2000, Evans2020}, conditions within dense, young star clusters \citep{Perets2012, Oh2016} and the Galactic Centre environment \citep{Rossi2017, Evans2022b, Evans2023, Verberne2024}.    




In this work we explore the frequency and strength of interactions between single stars and COBs in Milky Way GCs. Our aim in this work is to explore the degree to which these interactions have led to the ejection of single stars from Galactic GCs over the relatively recent Galactic past (500 Myr) and whether these interactions contribute significantly to the population of hypervelocity stars in the Galaxy. In Sec.~\ref{sec:methods} we outline how we assign realistic COB populations to Milky Way GCs, how we use a particle spray code to model the star + COB (S+COB) interactions, how we determine where escaped stars end up in the Galaxy today and how we obtain realistic mock photometric and astrometric observations of this escaped population. In Secs. \ref{sec:results:slow}, \ref{sec:results:fast} and \ref{sec:results:observability} we analyze these escapers, determining whether stars are ejected fast enough to be HVSs, and whether any should in princible be found in the \textit{Gaia} catalogue or should be detectable in the near future by the Vera C. Rubin Observatory's Legacy Survey of Space and Time (LSST). In Sec.~\ref{sec:discussion} we discuss the implications of these results and offer conclusions in Sec.~\ref{sec:conclusions}.

To clarify, in this work we use the term \textit{hypervelocity star} to refer to all fast stars ejected from Milky Way GCs with present day Galactocentric total velocities above $500 \; \mathrm{km \ s^{-1}}$. Such stars would be gravitationally unbound from the Galaxy as long as they are farther than $\sim$10 kpc from the Galactic Centre \citep{Deason2019} This is a somewhat looser definition of the term than elsewhere in the literature, where it is often reserved exclusively for unbound stars ejected from the Galactic Centre via the Hills mechanism. See \citet{Brown2015rev} for a review.

\section{Methodology} 
\label{sec:methods}

In this section we outline how we generate our mock populations of stars ejected from Milky Way GCs. We describe our GC sample, how we populate these GCs with COBs, how we model the S+COB interactions in these clusters, how stars which escape from GCs due to these interactions are propagated through the Galaxy, and how we apply mock observations to determine which of these stars are detectable by contemporary and near-future surveys.

\subsection{Globular Cluster Modelling}
\label{sec:CMC}

\begin{figure*}
    \centering
    \includegraphics[width=2\columnwidth]{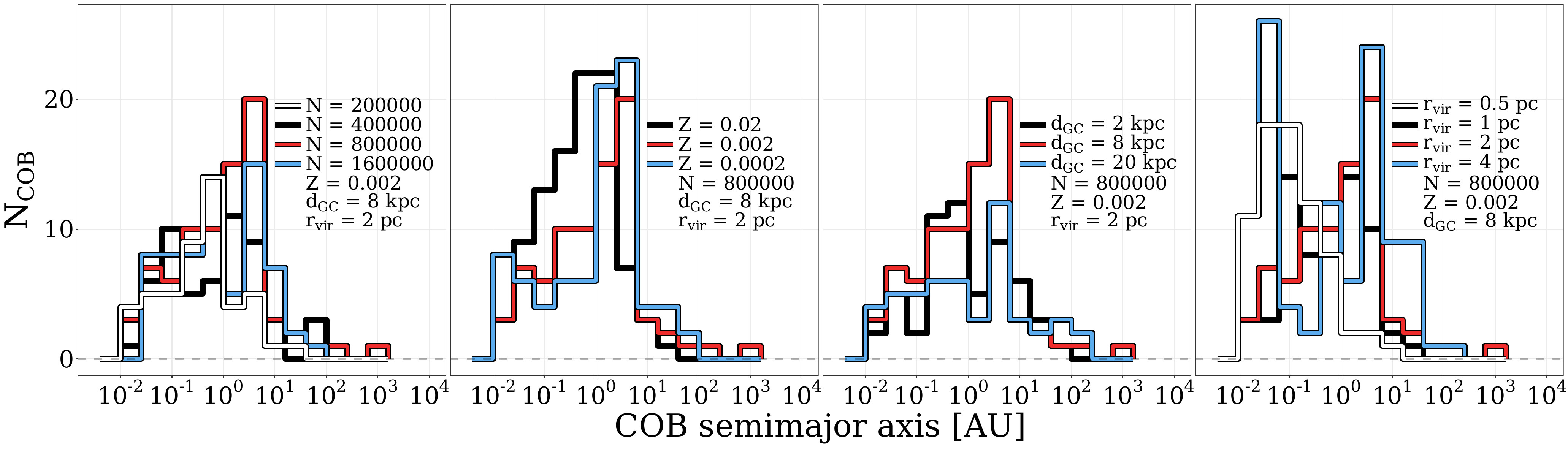}
    \includegraphics[width=2\columnwidth]{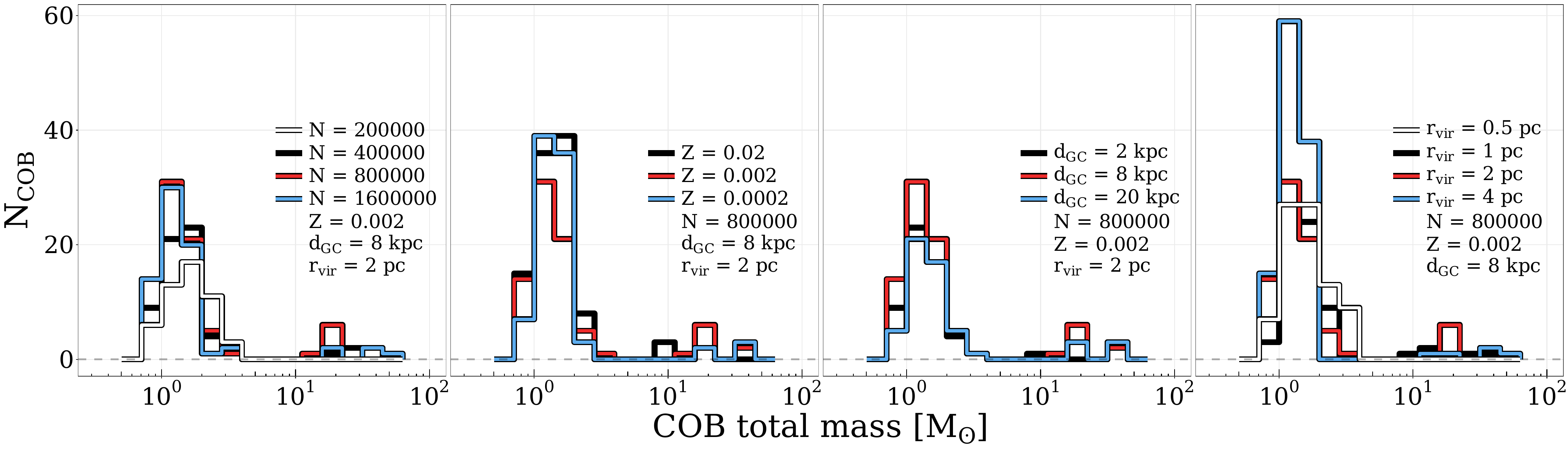}
    \caption{Distributions of separations (top row) and total binary masses (bottom row) of COB populations in the final timesteps of different \texttt{CMC} models and their dependence on the initial number of stars in the GC model (first column), their metallicity (second column), orbital semimajor axis with respect to the Galactic Centre (third column) and initial virial radius (fourth column). }
    \label{fig:BHmasses}
\end{figure*}

We use the outputs from the \texttt{Cluster Monte Carlo} (\texttt{CMC}) simulation suite to account for GC evolution. \texttt{CMC} is a H\'{e}non-type Monte Carlo code for simulating dense $N$-body systems \citep{Henon1971a, Henon1971b, Joshi2000, Joshi2001, Pattabiraman2013, Rodriguez2015, Kremer2020, Rodriguez2022}. \texttt{CMC} is able to model the structural evolution of a GC across cosmic time as well as the evolution of the objects which comprise it, taking into account a variety of physics including single and binary stellar evolution, compact object formation, strong and weak few-body interactions between objects, and stellar collisions. See \citet{Kremer2020} and \citet{Rodriguez2022} for more detailed explanations of the treatments of relevant physical processes. 

Following \citet{Kremer2020}, we use the set of 144 \texttt{CMC} models spanning a 4x4x3x3 grid in which the following parameters are varied independently: the initial total number of stars ($N=2\times10^{5}, 4\times10^{5}, 8\times10^5, 1.6\times10^6$), the metallicity ($Z = 0.0002, 0.002, 0.02$), the initial cluster virial radius ($r_{\rm vir} = 0.5, 1, 2, 4 \; \mathrm{pc}$) and the Galactocentric distance ($d_{\rm GC}=2, 8, 20 \; \mathrm{kpc}$). Each model assumes the density profile of the cluster follows a \citet{King1966} profile with an initial $W_0$ parameter of 5. Single stars have initial masses drawn from a \citet{Kroupa2001} initial mass function (IMF). 5\% of stellar systems are assumed to be initially in hard binaries, which is the typical fraction in GCs today \citep{Milone2012, Ji2015} and the overall fraction of hard binaries in GCs does not change significantly over a Hubble time \citep{Fregeau+2009}. Within these binaries, the mass of the primary star $m_1$ is also drawn from a \citet{Kroupa2001} IMF, the mass of the companion $m_2$ is drawn from a flat distribution in the mass ratio $0.1 \leq q \equiv m_1/m_2 \leq 1$, and the binary orbital period $P$ is drawn from a log-uniform distribution in the range $5(R_1+R_2) \leq a \leq a_{\rm hard}$, where $R_1$ and $R_2$ are the stellar radii, $a$ is the binary orbital separation, and $a_{\rm hard}$ is the hard/soft boundary \citep{Duquennoy1991}.

Each cluster in the grid is then evolved for a Hubble time. From each cluster simulation that did not disrupt before the present-day, we extract all retained binaries in which at least one member star is a compact object, i.e. a black hole (BH), neutron star (NS) or white dwarf (WD). There are eight configurations of these binaries, which we refer to as BHBH, BHNS, BHWD, NSNS, NSWD, WDWD, BHMS, NSMS, WDMS binaries, where ``MS" denotes a main sequence, non-compact star. To match these models with known Galactic GCs (see next subsection), the COB populations are extracted from snapshots spanning 9 Gyr to 14 Gyr. 

In Fig.~\ref{fig:BHmasses} we show the number of COBs retained in the final ($\sim$14 Gyr) timestep of the \texttt{CMC} models, their total mass and separation distribution, and how these change across the model grid. See \citet{Kremer2020} for a more detailed investigation into the compact object demographics within the \texttt{CMC} models. In general, as $N$ increases, the retained COB population increases slightly in number and favors wider binaries. The $N=2\times10^5$ models tend to be deficient in $M_{\rm tot}\gtrsim10 \; \mathrm{M_\odot}$ COBs which are the domain of BHBH and BHMS binaries. An increase in metallicity tends to lead toward tighter binaries but does not appear to greatly affect the total mass distribution. Interestingly, of the Galactocentric distances explored, the $d_{\rm GC}=8 \; \mathrm{kpc}$ models have a larger COB population than the models at larger and smaller distances. Finally, larger initial virial radii lead to wider COBs on average, and in particular the $r_{\rm vir} = 4 \; \mathrm{pc}$ models show a distinct bimodal distribution in COB separation. Notice as well that the $r_{\rm vir} = 0.5 \; \mathrm{pc}$ model show a lack of $M_{\rm tot}\gtrsim10 \; \mathrm{M_\odot}$ COBs. 

In terms of demographics, the most common COB type in the present day across all models are WDMS binaries, which typically constitute 65\% - 85\% of the total COB population. WDWD binaries are the second most common population at 10\% - 30\%. BHBH, BHMS and NSWD binaries each make up $\sim$a few percent of the population, while the remainder (BHNS, BHWD, NSNS and NSMS) are very rare at the present day.

\subsection{Matching Models to Milky Way Globular Clusters} \label{sec:methods:gcs}

With our evolved \texttt{CMC} model grid in hand, we can begin determining how many COBs are in known Milky Way GCs in the present day. Our catalogue of GCs is taken from \citet{Baumgardt2018}\footnote{\url{https://people.smp.uq.edu.au/HolgerBaumgardt/globular/}. Note that this is a live catalogue -- structural parameters may be updated periodically as new data becomes available. For posterity, the catalog used in this work is Version 4 of the catalog, updated March 2023.}, who provide structural parameters for 159 GCs along with 6D kinematic measurements from \citet{Baumgardt2019, Vasiliev2021, Baumgardt2021}.  The GC catalogue of \citet{Harris2010} provides metallicity measurements for 149 of these clusters; we discard the remainder.

We choose an appropriate \texttt{CMC} model for each GC following an approach similar to \citet{Rui2021} and \citet{Cabrera2023}. Our approach is as follows. We have 144 \texttt{CMC} models in a grid of initial population size, virial radius, metallicity and Galacocentric distance. For each of these models we have outputs for an average of $\sim$16 timesteps spanning 9 Gyr and 14 Gyr, for a total of 2248 model outputs. For each Galactic GC, we first identify the subset of model outputs closest to it in standardized (i.e. rescaled such that the mean is zero and the dispersion is 1) log$Z - d_{\rm GC}$ space, since these GC parameters do not vary with time in the models. For each output timestep for each of these models in the subset, we determine their positions in standardized log$M - r_{\rm c}/r_{\rm hl}$ space, where $M$ is the total mass, and $r_{\rm c}$ and $r_{\rm hl}$ are the core radius and half-light radius, each of which \textit{do} vary in time. A model is chosen at random from among the \textit{three} closest output timesteps to the Galactic GC in this space and its COB population is assigned to the Galactic GC. Since our predictions for the escaped S+COB population in the Galaxy later on will be averaged over fifty repeated iterations, multiple different COB populations will therefore be assigned to the Galactic GC and our overall predictions will be less susceptible to stochasticity and model variations. In Sec. \ref{sec:discussion:model} we discuss the impact of our GC-matching scheme on results.

\begin{figure*}
    \centering
    \includegraphics[width=\columnwidth]{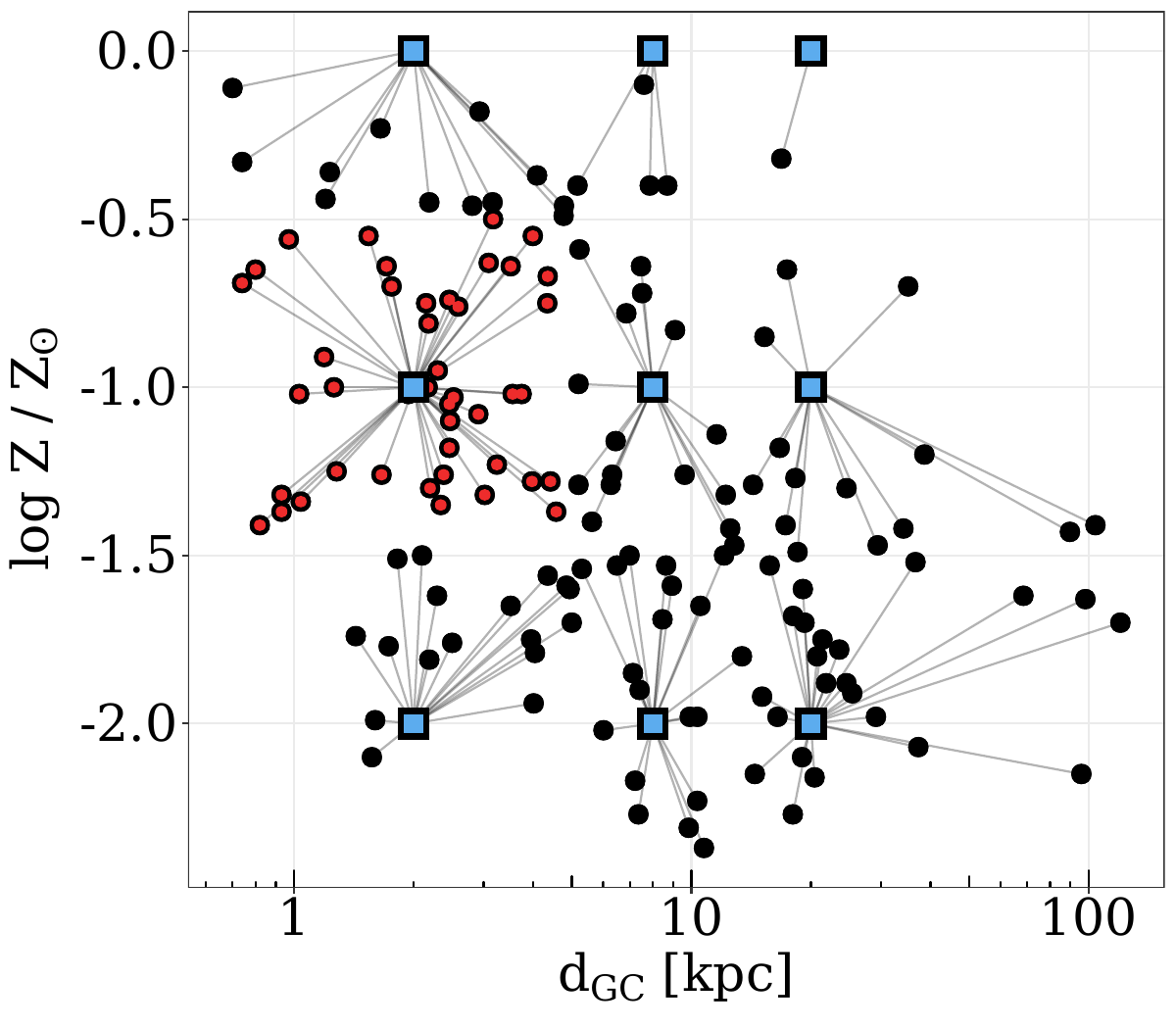}
    \includegraphics[width=\columnwidth]{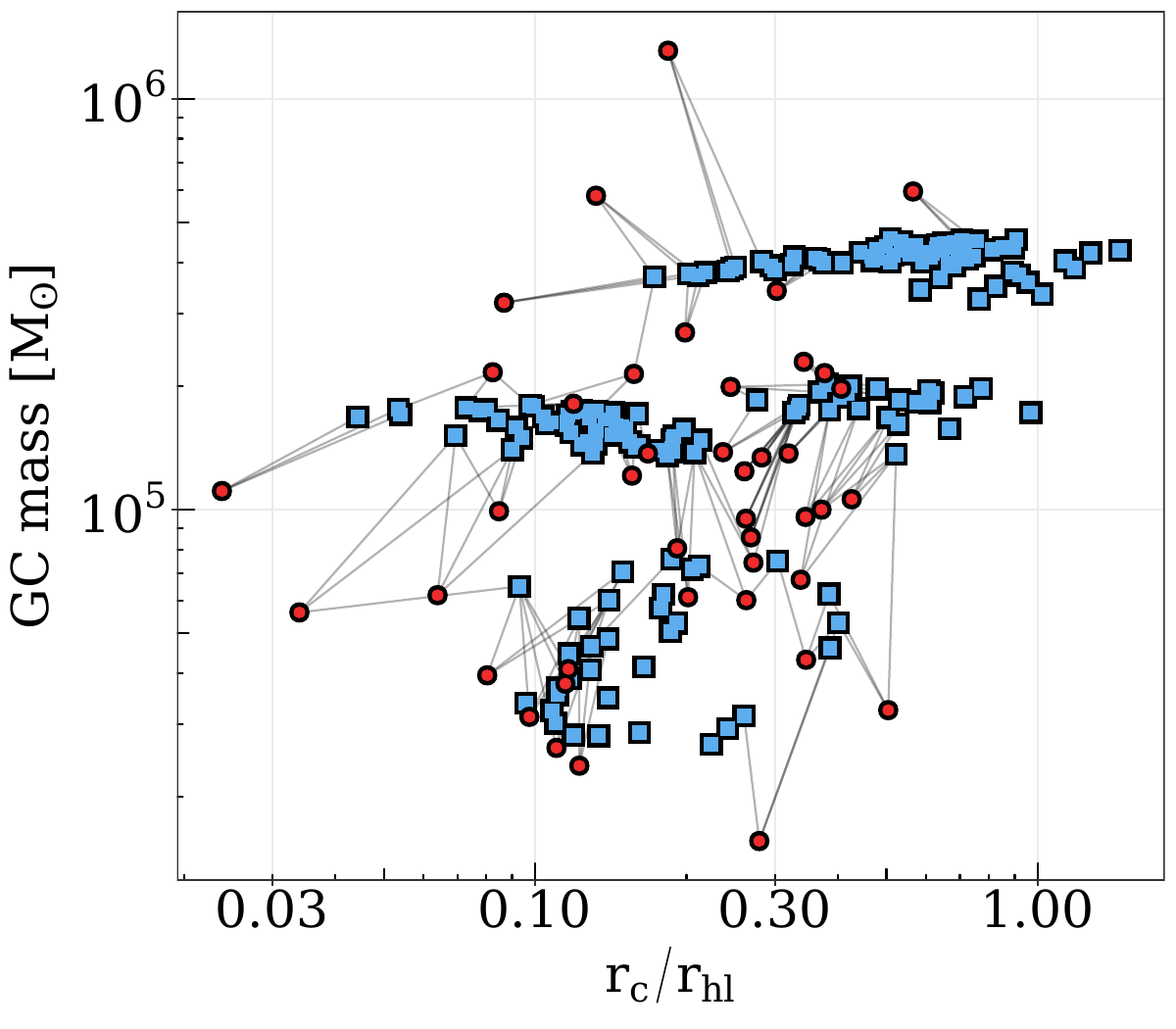}
    \caption{\textit{Left:} The Galactocentric distances and metallicities of the \texttt{CMC} models we use in this work (blue squares) and our Milky Way GC sample (black and red points). Line segments connect each Milky Way GC to the closest \texttt{CMC} model subset. \textit{Right:} The total mass and ratio of core radius to half light radius among the Milky Way GCs closest to log$[Z/Z_\odot]=-1, \;d_{\rm GC}=2$ kpc (shown in red in the left panel). The blue squares show the masses and $r_{\rm c}/r_{\rm hl}$ ratios for all timesteps in the subset of \texttt{CMC} models with log$[Z/Z_\odot]=-1$ and $d_{\rm GC}=2$ kpc . Line segments connect each Milky Way GC to its three closest \texttt{CMC} outputs in this space.}
    \label{fig:matching}
\end{figure*}

We illustrate this matching in Fig.~\ref{fig:matching}, where we show how the Milky Way GC and \texttt{CMC} models populate the log$Z-d_{\rm GC}$ (left) and $\log M - r_{\rm c}/r_{\rm hl}$ (right) spaces. In the left-hand panel, each line segment connects a Milky Way GC to its appropriate set of closest \texttt{CMC} models in log$Z - d_{\rm GC}$ space. In the right-hand panel, we show how the Milky Way GCs closest to one example point in log$Z  - d_{\rm GC}$ space (log$[Z/Z_\odot]= -1, \;d_{\rm GC}=2$ kpc) populate $\log M - r_{\rm c}/r_{\rm hl}$ space. Line segments connect each of these GCs to the three closest \texttt{CMC} model timesteps in this space from among the models with log$[Z/Z_\odot] = -1$ and $d_{\rm GC}=2$ kpc. Matching real GCs to the most similar model GC timesteps in this way introduces some uncertainty into our modelling, since minor changes in the cluster parameters can lead to changes in the COB population. Simulations tailored to each Galactic GC would be preferred, but the advantage of using the existing \texttt{CMC} simulations is that these models are already available and widely used.

\subsection{Modelling S+COB encounters}
\label{sec:methods:encounters}

\subsubsection{Estimating the interaction rate}\label{sec:rate}

With our catalog of Galactic GCs and the COBs they contain, we determine the frequency with which S-COB interactions occur in the cluster cores. Under the assumption that both the cluster core properties and COB demographics remain relatively unchanged in the recent past, we calculate the S+COB interaction rates for each cluster in the present day and assume this rate was the same 500 Myr ago. We impose this maximum flight time of 500 Myr because any S+COB HVSs which escape earlier than this will be extremely distant from the Milky Way and therefore very difficult to detect.

Following \citet{Leonard1989} the rate $\Gamma$ at which bodies of species $x$ experience a strong interaction with species $y$ in a many-body system is 

\begin{equation} \label{eq:Gamma}
    \Gamma = \int \gamma d^3 r = \int 0.5 n_x n_y \langle\sigma_{x+y} v_{\rm rel} \rangle d^3r \; \text{,}
\end{equation}
where $\gamma$ is the interaction rate per unit volume, $n_x$ and $n_y$ are the species number densities, $v_{\rm rel}$ is the relative velocity between the species at pericenter, and $\sigma_{x+y}$ is the gravitationally focused interaction cross section expressed as
\begin{equation} \label{eq:sigma}
    \sigma_{x+y} = \pi p^2 [1 + \frac{2G(m_x + m_y)}{pv_{\rm rel}^2}] \; \text{,}
\end{equation}
where $p$ is the size of species $y$, $m_x$ and $m_y$ are the masses of species $x$ and $y$ respectively. The quantities in the integrand of Eq.~(\ref{eq:Gamma}) are roughly constant throughout GC cores in the present day. Recognizing this, and assuming species $x$ is single stars (S) and species $y$ is each type of COB (i.e. BHBH, BHNS, BHWD, NSNS, NSWD, WDWD, BHMS, NSMS, WDMS),

\begin{equation}
    \Gamma_{S+i} = 0.5n_{\rm S} n_{\rm i} V \langle\sigma_{\rm S+i} v_{\rm rel} \rangle \; \text{,}
\end{equation}
where $i$ is the COB type and $V$ is the cluster core volume. Assuming a Maxwellian distribution, $\langle\sigma_{S+i} v_{\rm rel} \rangle = 2v_{\rm rms}\sigma_{S+i}(v_{\rm rms})$, where $v_{\rm rms}$ is the root mean squared relative velocity and $\sigma_{S+i}(v_{\rm rms})$ is Eq.~(\ref{eq:sigma}) evaluated at $v_{\rm rms}$. Ignoring the first term of Eq.~(\ref{eq:sigma}) since gravitational focusing dominates, $\Gamma_{S+i}$ becomes
\begin{equation}
    \Gamma_{S+i} = 2 \pi n_{\rm s} n_{\rm i} V G \langle p \rangle \frac{\langle m_{\rm S} \rangle + \langle m_{\rm i}\rangle }{v_{\rm rms}} \; \text{.}
\end{equation}

Let us say $n_{\rm S} = (1 - f_{\rm mult}) \rho_c / \langle m_{\rm } \rangle$, where $f_{\rm mult}$ is the fraction of stellar systems in the cluster with one or more companions, $\rho_c$ is the cluster core density and $\langle m_{\rm } \rangle$ is the average mass of bodies in the cluster core. Let us also say $n_{\rm i} = f_{\rm i} \rho_c / \langle m_{\rm } \rangle$, where $f_{\rm i}$ is the fraction over \textit{all} stellar systems which are the COB population $i$, taken from the \texttt{CMC} model output matched to the GC. The interaction rate is then

\begin{figure}
    \centering
    \includegraphics[width=\columnwidth]{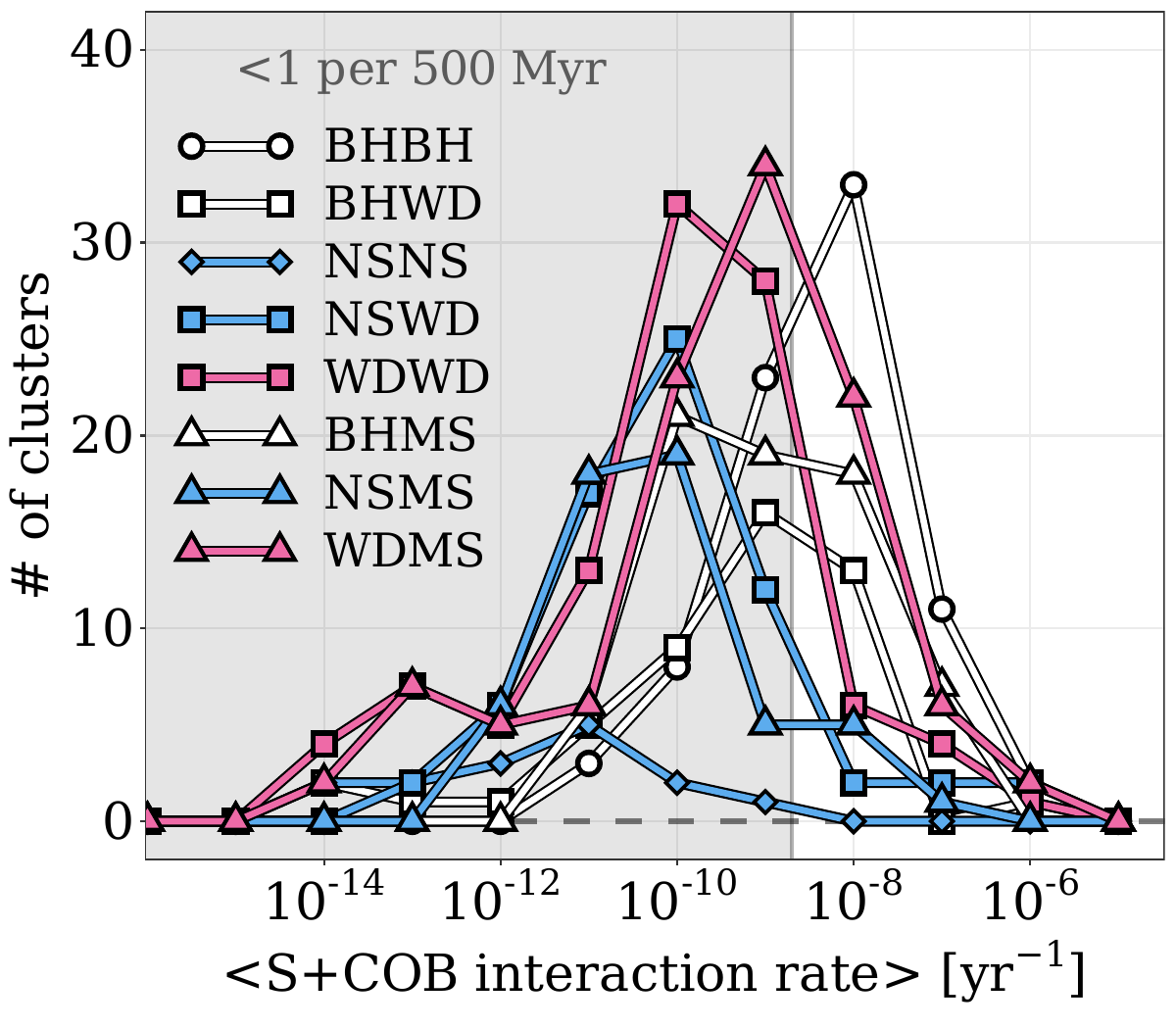}
    \caption{A histogram showing the mean S+COB interaction rate in Galactic GCs, separated by COB type. The color of the line denotes the type of the primary of the COB and the symbol denotes the type of the secondary. The shaded region marks where $\leq$ 1 interaction per cluster per 500 Myr occurs. BHNS binaries are not shown because no Galactic GC is matched to a model which includes retained BHNS binaries.}
    \label{fig:rates}
\end{figure}

\begin{equation} \label{eq:rate}
    \Gamma_{S+i} = \frac{8}{3}\pi^2 G \langle p \rangle (1-f_{\rm mult}) f_{\rm i} r_{\rm c}^3 \frac{\rho_{\rm c}^2}{\langle m_{\rm } \rangle ^2} \frac{\langle m_{\rm S} \rangle  + \langle m_{\rm i}\rangle}{v_{\rm rms}} \; \text{.} \\
\end{equation}


We assign a rate $\Gamma_{S+i}$ for each COB type for each cluster assuming that $\langle p \rangle$ is the mean semimajor axis  of the $i$'th COB population in that cluster. $f_{\rm mult}$ varies across different GCs and depends on the masses of the primary and companion(s), so for simplicity we assume a fixed multiplicity fraction of $f_{\rm mult} = 0.05$ across all GC \citep{Milone2012, Ji2015, Zhang2024}. We take $f_{\rm i}$ and $\langle m_{\rm i} \rangle$ from the matched \texttt{CMC} model, assume $\langle m \rangle=\langle m_s \rangle \approx 0.35 \; \mathrm{M_\odot}$ is the typical mass of a single star for a \citet{Kroupa2001} initial stellar mass function evolved for the age of the GC using \texttt{McLuster} \citep{Kupper2011}, and take $r_{\rm c}$, $\rho_{\rm c}$ and $v_{\rm rms}$ directly from our Milky Way GC catalogue. Recall that since one of three closest \texttt{CMC} model is chosen at random to be matched to a GC, the estimated interaction rate will change from iteration to iteration. In Fig.~\ref{fig:rates} we show histograms of the mean $\Gamma_{S+i}$ for each cluster separated by COB type. The interaction rate per cluster, when non-zero, spans $\sim10^{-14}-10^{-6} \; \mathrm{yr^{-1}}$. The COB types with the largest overall interactions rates are WDMS, BHBH, NSWD and BHMS, whose total rates across all clusters are $3.0 \times 10^{-6} \; \mathrm{yr^{-1}}$, $2.0 \times 10^{-6} \; \mathrm{yr^{-1}}$, $1.2 \times 10^{-6} \; \mathrm{yr^{-1}}$ and $1.1 \times 10^{-6} \; \mathrm{yr^{-1}}$, respectively. The COB types with the smallest interaction rate is BHNS binaries, which is left off this plot entirely since over fifty repeated iterations, none of our Galactic GCs are ever matched to a \texttt{CMC} model with a retained population of BHNS binaries. The dependence of these interaction rates on the GC properties is discussed further below.

\subsubsection{Simulating encounters with \texttt{Corespray}} \label{sec:corespray}

We use the particle spray code \texttt{Corespray}\footnote{\url{https://github.com/webbjj/corespray}}\citep{Grondin2023} to model the S+COB interactions. \texttt{Corespray} samples the outcomes of three-body interactions for any Milky Way GC using the theoretical three-body framework outlined in \cite{Valtonen2006}. \texttt{Corespray} offers a variety of adjustable input parameters, including structural and orbital information of specific GCs, initial encounter separations, system masses and binding energies. Here, we briefly describe our modeling of three-body encounters using \texttt{Corespray}, but refer the reader to \citet{Grondin2023, Grondin2024GEMS} for more detailed information. 


Using the interaction rates for each GC and each COB type computed in Sec. \ref{sec:rate}, we consider a total of $N_i = \Gamma_{S+i} \times 500 \; \text{Myr}$ interactions in each cluster and each COB type in our catalogue, where, for reminder, $i$=[BHBH, BHNS, BHWD, NSNS, NSWD, WDWD, BHMS, NSMS, WDMS] are the eight different COB types we explore. We assume these interactions occur at times uniformly distributed between now and 500 Myr in the past. 

For each interaction, we first select a COB at random from among the $i$'th COB type in the appropriate \texttt{CMC} model output selected following Sec. \ref{sec:CMC}. The mass of the single MS star interacting with this COB is drawn from a \citet{Kroupa2001} initial mass function which has been evolved for 12 Gyr using \texttt{McLuster} \citep{Kupper2011}.
We randomly sample an encounter radius between the semi-major axis of the COB and the mean separation of stars in each cluster's core. This ensures that we span a wide range of encounter radii. The gravitational potential of each GC is modelled using a King profile \citep{King1966}, whose central parameter $W_0$ is determined from its concentration $c\equiv r_{\rm t}/r_{\rm c}$ using the \texttt{Python} package \texttt{clustertools} \citep{Webb2022}, where $r_{\rm t}$ and $r_{\rm c}$ are the cluster tidal radius and core radius, respectively. The core radius for each cluster is taken directly from the \citet{Baumgardt2018} catalogue, while for $r_{\rm t}$ we follow \citet{Grondin2023} and use Eq. of \cite{Webb2013} to compute $r_{\rm t}$ at apogalacticon for each cluster. 

\begin{figure}
    \centering
    \includegraphics[width=\columnwidth]{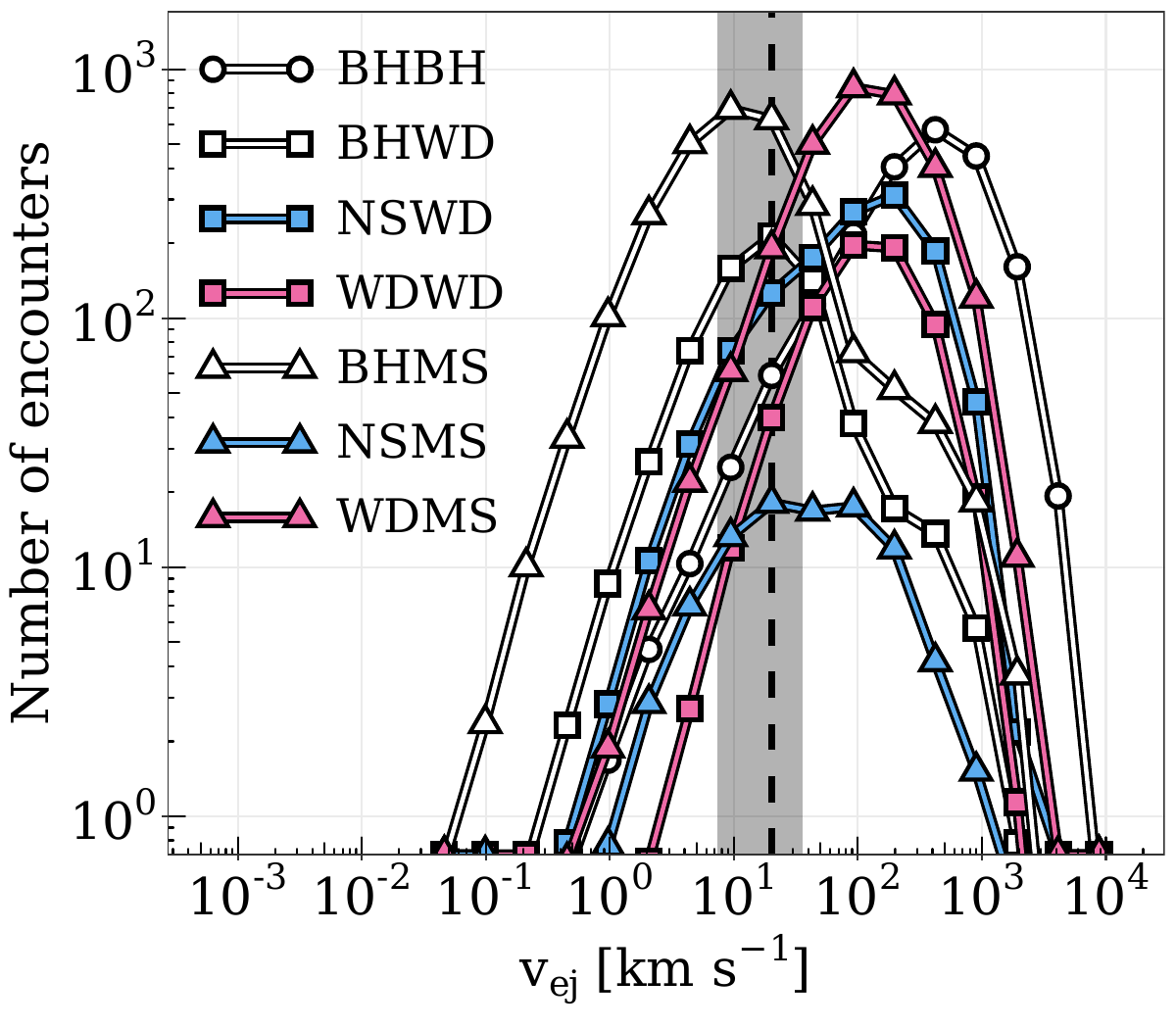}
    \caption{The distribution of ejection velocities over all S+COB encounters over the last 500 Myr in Milky Way GCs, sorted by COB type. The color of the line denotes the type of the primary of the COB and the symbol denotes the type of the secondary. The vertical dashed line denotes the median cluster escape velocity in our GC catalogue and the shaded region spans the $\pm1\sigma$ spread. The BHNS and NSNS distributions are not shown because $<$1 S+BHNS or S+NSNS interactions have occured on average in the last 500 Myr in Galactic GCs.}
    \label{fig:vej}
\end{figure}

Using the COB and single star masses and the encounter distance, \texttt{Corespray} samples the results of the encounter and determines the outcome of the interaction, including the ejection velocity $v_{\rm ej}$ of the single star in the centre of mass frame of the system, and the recoil velocity of the binary. In Fig.~\ref{fig:vej} we show the ejection velocity distribution for each COB type, stacked across all clusters and averaged over all iterations. The dashed line and shaded region show typical escape velocities of our GC clusters. Interestingly, each COB type shown is capable of interactions which can eject single stars at $\sim$several hundred $\mathrm{km \ s^{-1}}$, though for many COB types these ejections are quite rare. Interactions with BHBH binaries appear to eject stars with the largest average ejection velocity and contribute most strongly at the high-velocity end of the distribution. WDMS binaries, also eject a fair number of stars at high velocities. 

\begin{figure*}
    \centering
    \includegraphics[width=2\columnwidth]{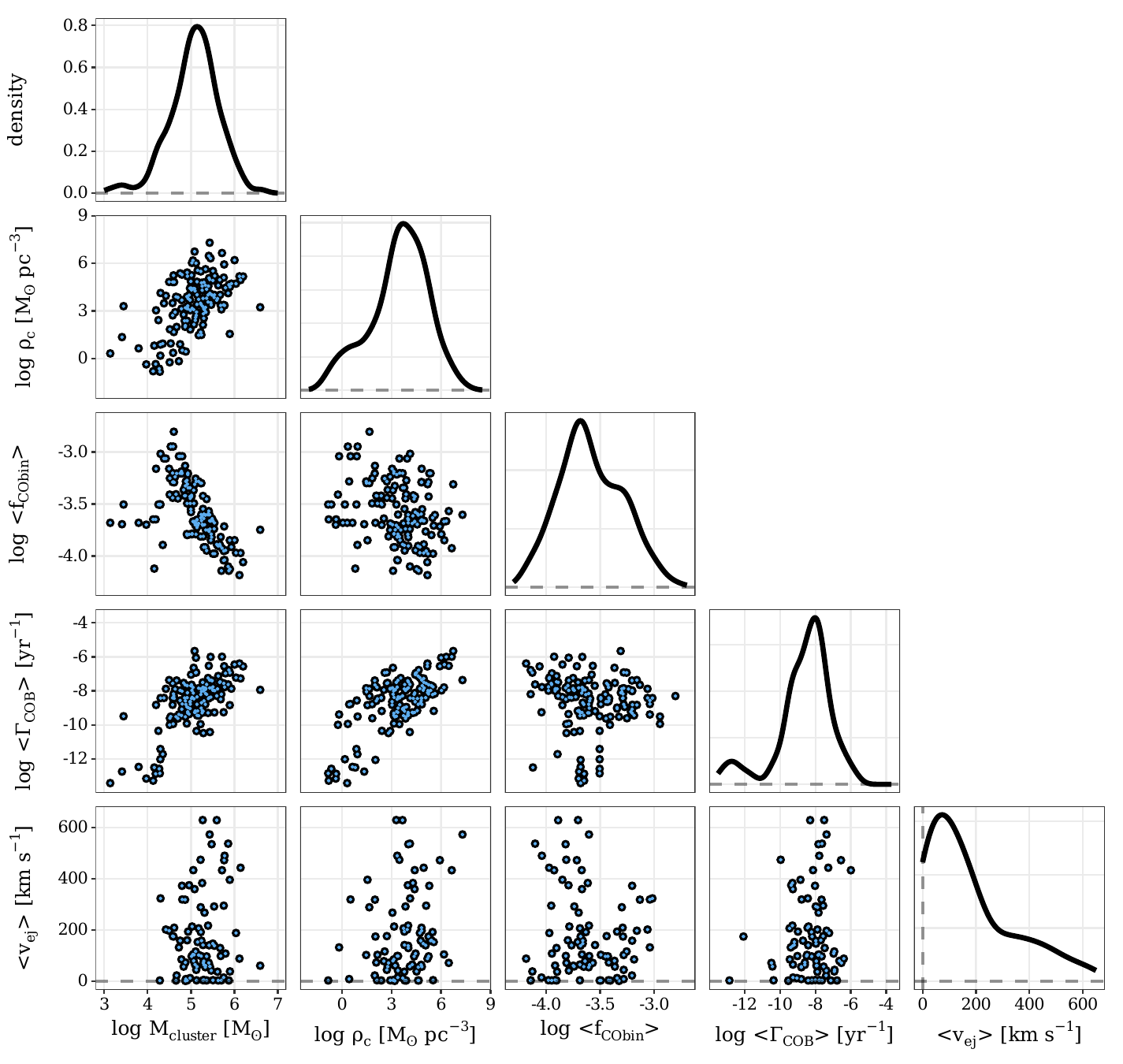}
    \caption{Among the 149 globular clusters in our sample, the distributions of and among the total cluster mass, stellar density in the core, the mean S+COB interaction rate, the mean fraction of stellar systems which are COBs, and the median ejection velocity from a S+COB interaction.} 
    \label{fig:StairStep_cluster}
\end{figure*}

Secs. \ref{sec:methods:gcs}-\ref{sec:methods:encounters} are summarized in Fig.~\ref{fig:StairStep_cluster}, which shows how the mean S+COB total interaction rate over fifty iterations, mean total fraction $f_{\rm COB} \equiv \sum_i f_i$ of systems which are COBs, and the median ejection velocity $\langle v_{\rm ej} \rangle $ for each cluster correlate with each other and with the central density and total mass of each GC. The interaction rate is correlated with the cluster central density (unsurprising since it factors directly into its calculation, see Eq.~\ref{eq:rate}) and note that it spans a wide range from $<10^{-12} \; \mathrm{yr^{-1}}$  to $10^{-5} \; \mathrm{yr^{-1}}$. The typical median ejection velocity from S+COB interactions in a GC is slightly slower than $200 \, \mathrm{km \ s^{-1}}$, still significantly above the cluster escape velocity (see Fig.~\ref{fig:vej}). For some clusters the typical ejection velocity can be as high as $\sim 500 \; \mathrm{km \ s^{-1}}$, but these clusters have middling interaction rates. The COB system fraction is typically 0.1-1 per thousand, and note that (perhaps unintuitively) this fraction is uncorrelated with the interaction rate and anti-correlated with the cluster total mass. 


It is worth pointing out which clusters have the largest S+COB interaction rates and asking which properties of these clusters lead to the large rates. Out of the 149 clusters, 59 have likely not played host to a single S+COB interaction across all COB types over the last 500 Myr, while 25\% of all S+COB interactions are in NGC 7099. The top nine GCs in terms of S+COB interaction rate (NGC 7099, NGC 7078, NGC 1851, NGC 6287, NGC 6388, Liller 1, NGC 6541, NGC 6266, and NGC 6715, in that order) together contribute 75\% of all interactions. Referring back to Eq.~(\ref{eq:rate}), clusters can have a high S+COB interaction rate by having favorable cluster properties (large, dense cores with small central velocity dispersions). When the clusters in our sample are sorted by $r_c^3\rho_c^2 v_{\rm rms}^{-1}$, these nine clusters are all within the top ten, though not in the same order. An interesting case is NGC 5694, which has the seventh-highest $r_c^3\rho_c^2 v_{\rm rms}^{-1}$ but is only rank 21 in its total S+COB ejection rate. This drop-off is due to the other main factor impacting the ejection rate -- containing a COB population favorable to a large interaction cross section (i.e. massive binaries with significant separations). It is also worth mentioning that none of these top nine most-ejecting clusters are ever matched to the same \texttt{CMC} model output, so our results are not particularly sensitive to the behavior of a single \texttt{CMC} model. 


\subsection{The orbital evolution of escaping stars} \label{sec:methods:prop}

Stars ejected faster than the escape velocity of their host cluster ($\sim$several $\mathrm{km \ s^{-1}}$ for typical clusters) will leave their host GC and travel through the Galaxy. Stars escaping from GCs following S+COB encounters are propagated through the Milky Way from their ejection time to the present day using the \texttt{Python} package \texttt{galpy}\footnote{\url{https://github.com/jobovy/galpy}}. In each iteration, we assign present-day kinematics to each GC by sampling its heliocentric distance, heliocentric radial velocity and proper motions from the measurements and uncertainties provided in the \citet{Baumgardt2018} catalogue, assuming Gaussian errors and taking into account the correlation between proper motion components (no uncertainty is assigned to the cluster sky positions). From these kinematics, we determine the orbital histories of each GC by back-propagating them through the Galactic potential \texttt{MWPotential2014} \citep{Bovy2015}. Stars ejected at an escape time $t_{\rm esc}$ ago are initialized at the central location of their host cluster $t_{\rm esc}$ in the past with their ejection velocity $v_{\rm esc}$ in a direction that is a random orientation in the cluster frame, such that its initial velocity in the Galactocentric frame is the vector addition of $v_{\rm esc}$ and the orbital velocity of its parent cluster $t_{\rm esc}$ ago. Each star is then integrated to the present day through a Galactic potential which consists of the sum of \texttt{MWPotential2014} and the moving King potential of its parent cluster using a 8(5,3) \citet{Dormand1980} integrator with a timestep of 0.1 Myr. \citet{Grondin2024GEMS} recently showed that choice of potential (either time-independent and time-dependent) has little effect on the present-day distributions of escaped stars from three-body encounters in GC cores. We have confirmed that this holds true for stars escaping via S+COB interactions as well -- the results presented here remain unchanged if the stars are propagated through other reasonable and widely-used Milky Way potentials, e.g., as described in \citet{McMillan2017}.

\subsection{Mock photometry of escaped stars} \label{sec:methods:phot}
\begin{figure*}
    \centering
    \includegraphics[width=2\columnwidth]{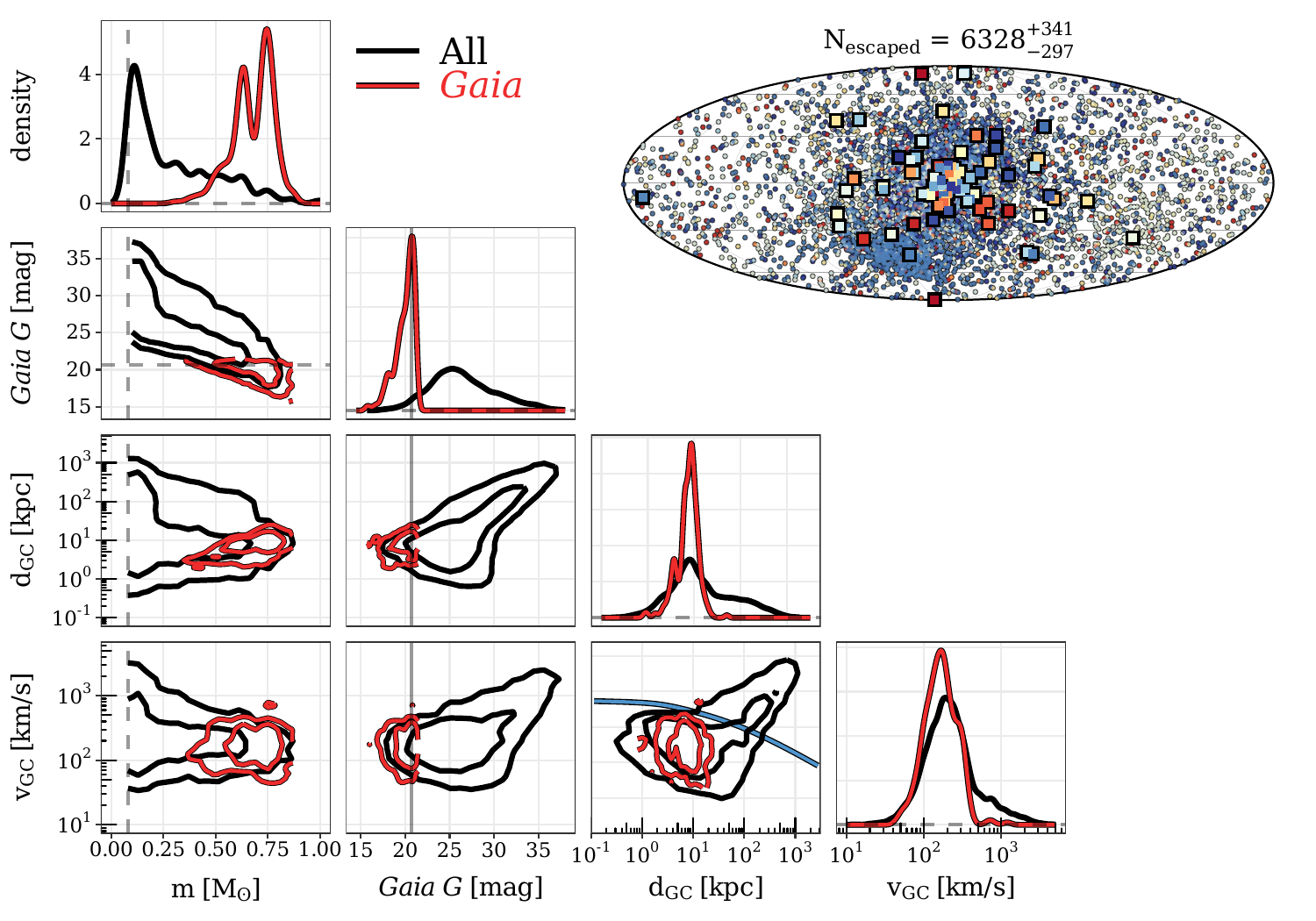}
    \caption{For all stars which escaped in the last 500 Myr from S+COB interactions in GC cores, the distributions of their stellar mass, \textit{Gaia G}-band apparent magnitude, Galactocentric distance, and Galactocentric total velocity. The inner and outer contours enclose 68\% and 95\% of the population of all stars (black curves) and of only those stars detectable by \textit{Gaia} (red curves). The vertical solid line in the second column shows the \textit{G}=20.7 faint-end magnitude limit of the \textit{Gaia} source catalog. The sky distribution of all $6330_{-300}^{+340}$ escaped stars in Galactic coordinates is shown in the top right, where each star shares a color with the GC (squares) from which it originates.} 
    \label{fig:StairStep_all}
\end{figure*}

After generating and propagating the simulated population of escaped stars, we estimate the mock apparent magnitudes of each star in the \textit{Gaia} and LSST photometric bands. This is useful in determining how many and which of the escaped stars could in principle be detectable in the present day and near future.

We start by determining the visual dust attenuation at each star's distance and sky position using the \texttt{combined15} dust map of \citet{Bovy2016}\footnote{\url{https://github.com/jobovy/mwdust}}. From the mass, metallicity and extinction of each mock star, we determine its magnitudes in the \textit{Gaia} $G$ and $G_{RP}$ band, the Johnsons-Cousins $V$ and $I_{\rm c}$ bands \citep{Bessell1990} and the LSST $g$ and $r$ bands \citep{DES2018} using the \texttt{Python} package \texttt{BRUTUS}\footnote{\url{https://github.com/joshspeagle/brutus}}, which uses Bayesian inference to generate photometry based on a grid of stellar models. In this case we employ the MESA Isochrone and Stellar Tracks, or \texttt{MIST} \citep{Dotter2016, Choi2016} models\footnote{\url{https://waps.cfa.harvard.edu/MIST/}}. In this estimation we assume all escaped stars share the metallicity of their parent GC. For the purpose of calculation, all stars are assigned the same age as the \texttt{CMC} output timestep matched with their parent GC, though the precise assumed age has very little impact on their inferred magnitudes. 

We use the tools provided by the GaiaUnlimited\footnote{\url{https://gaiaunlimited.readthedocs.io/}} project to determine which mock stars should be detectable by \textit{Gaia} and would be included in its third and most recent data release (DR3). We begin by querying the provided DR3 empirical selection function \citep{CantatGaudin2022}, which gives the probability $P(\rm DR3)$ that a star appears in the source catalogue based on its sky position and $G$-band magnitude. Next, following \citet{CastroGinard2023} Appendix D, we construct the selection function for the \textit{Gaia} subsample with measured five-parameter astrometry (position, parallax, proper motion) as well as a renormalized unit weight error (RUWE) smaller than 1.4. Querying this subsample selection function with position, \textit{Gaia G}-band magnitude and $G-G_{RP}$ color yields the probability $P({\rm ast}|\rm DR3)$ that a star in the DR3 source catalogue also has a well-behaved astrometric solution. The total probability for a star appearing in \textit{Gaia} DR3 with a well-behaved astrometric solution is then $P(\mathrm{ast}) = P(\mathrm{ast}|\mathrm{DR3}) \cdot P(\rm DR3)$. We generate a random number $\epsilon_{\rm ast}$ uniformly between 0 and 1 for each star and designate it as \textit{Gaia}-detectable if $\epsilon_{\rm ast} \leq P{\rm (ast)}$. We similarly query the GaiaUnlimited-provided \textit{Gaia} DR3 radial velocity subsample selection function to determine the probability $P(\mathrm{RV}) = P(\mathrm{RV}|\mathrm{DR3}) \cdot P(\rm DR3)$ that each mock star would appear in the \textit{Gaia} DR3 radial velocity catalogue. We designate stars as detectable in the \textit{Gaia} DR3 radial velocity catalogue if a random number $\epsilon_{\rm RV}$ is less than $P(\rm RV)$.

After determining which escaped stars should in principle be included in the \textit{Gaia} DR3 astrometric and radial velocity catalogues, we estimate \textit{Gaia} measurement uncertainties as well. For the astrometric uncertainties we use the \textit{Gaia} (E)DR3 astrometric spread function of \citet{Everall2021cogiv}\footnote{\url{https://github.com/gaiaverse/scanninglaw}}, which provides the full astrometric covariance matrix of a source depending on its sky position and \textit{G}-band magnitude. We estimate radial velocity uncertainties using the \texttt{Python} package \texttt{PyGaia}\footnote{\url{https://github.com/agabrown/PyGaia}} based on the pre-launch predicted performance of the satellite.

Our selection of stars detectable by the upcoming LSST is more straightforward in comparison. LSST will observe the entire southern equatorial hemisphere along with regions as north as $\delta\approx+33.5^\circ$ in the eastern sky and as north as $\delta\approx+15^\circ$ in the west. We designate LSST-detectable mock stars as those which fall within the area of the most recent (v4.3.1) survey baseline\footnote{see \url{https://survey-strategy.lsst.io/baseline/index.html}} and have an apparent magnitude in the $r$ band between the bright-end limit of 16 and the single-visit faint-end limit of $24.5$ \citep{Marshall2017, Ivezic2019}.

Finally, we remove all stars which lie within 1.5$\times$ the tidal radius of any GC in our sample on the sky. This cuts out all stars ejected very recently and/or with velocities entirely along the line of sight, or stars which by chance are currently overlapping another GC on the sky. Such escaped stars would be difficult to identify in practice.
 
It is important to caution that a star being detectable by \textit{Gaia} and/or LSST does not at all imply that it will be easily \textit{identifiable} as a star originally escaped from a GC. Without detailed kinematic and/or chemical information, cluster stars ejected at low-to-modest escape velocities will be difficult to distinguish from old stellar populations in the Milky Way bulge and stellar halo. In Sec.~\ref{sec:results:observability} we delve further into the observational realities of identifying these stars. Furthermore, a forthcoming work by Battson et al. (in prep) will perform a more in-depth observational investigation into \textit{Gaia} stars in the immediate vicinity around globular clusters with a velocity above the central escape velocity of the cluster. This is in order to detect recent ejections from the cluster which may come from a variety of escape mechanisms, including those involving compact objects.

\section{Escaping stars ejected by S+COB encounters in globular clusters} 
\label{sec:results:slow}


\begin{figure*}
    \centering
    \includegraphics[width=2\columnwidth]{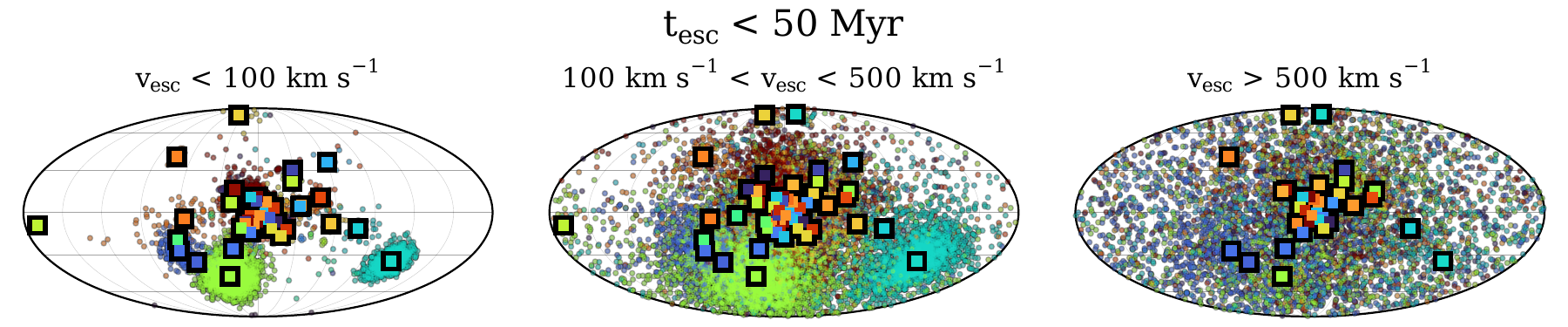}
    \includegraphics[width=\columnwidth]{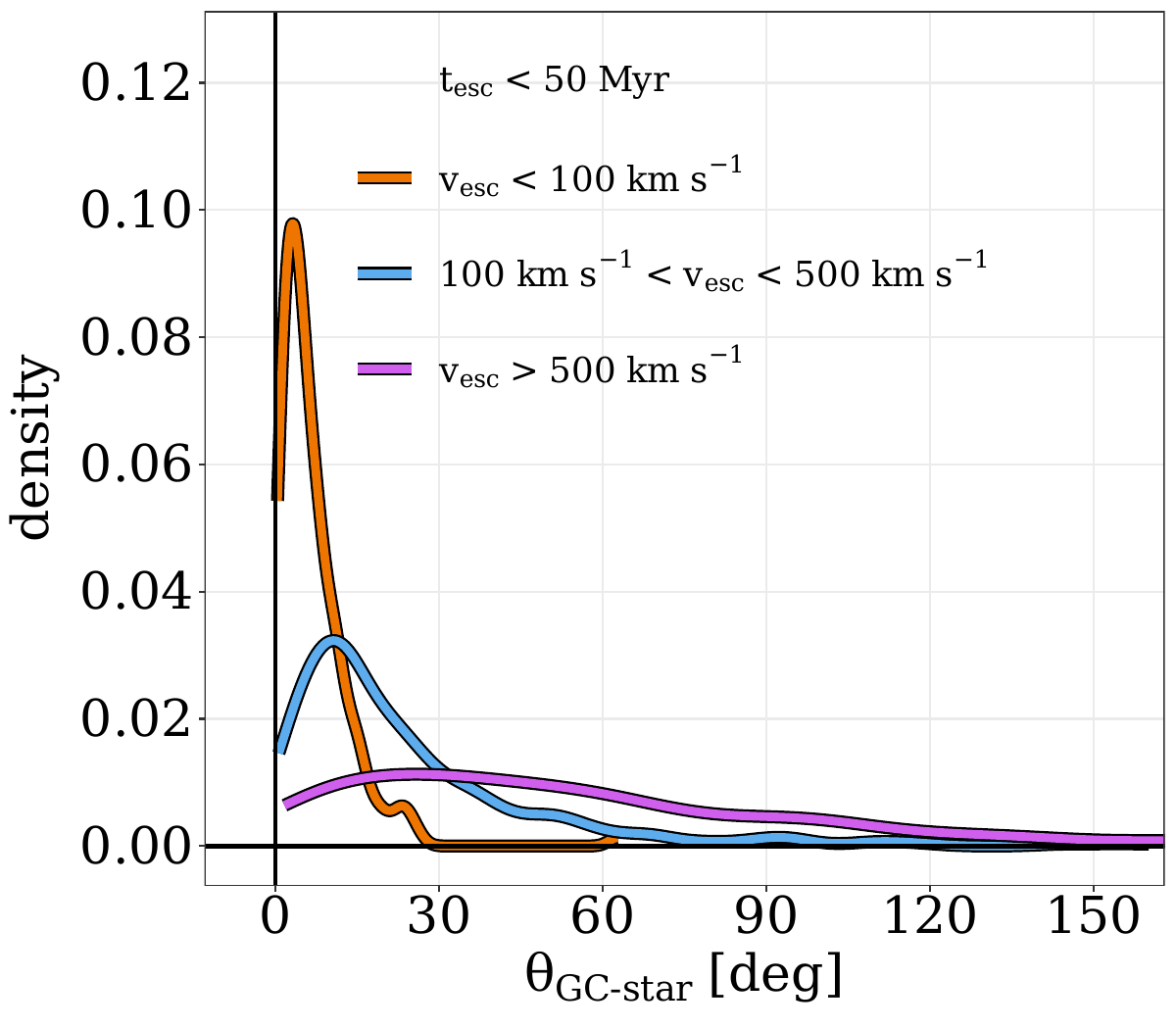}
    \includegraphics[width=\columnwidth]{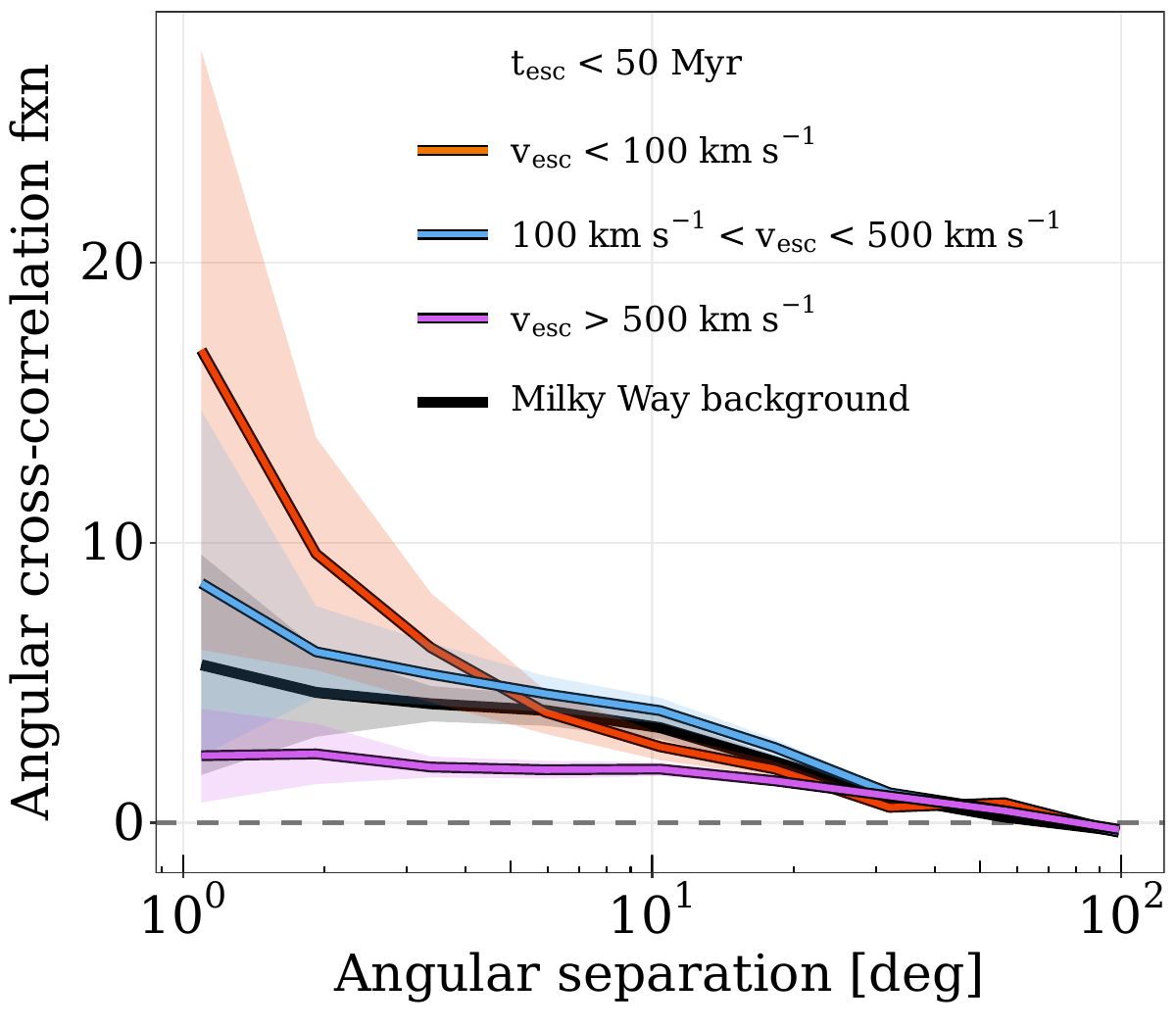}

    \caption{\textit{Top:} The same as the upper right panel of Fig.~\ref{fig:StairStep_all} but only for stars which have recently ($t_{\rm esc} < 50 \; \mathrm{Myr}$) escaped GCs from S+COB interactions. Distributions for three different bins of ejection velocity are shown. Clusters (squares) are only shown in a panel if 100 or more stars which have escaped from them are in the bin. \textit{Lower left:} The angle between the present day positions of recently escaped S+COB stars and their natal GC. \textit{Lower right}: The two-point angular cross-correlation function between GCs and recently escaped S+COB stars, i.e. the excess probability (over a random uniform distribution) to find an S+COB star separated from any GC in our sample at a particular angular scale. Shaded regions show the  1$\sigma$ variation over 100 iterations. The black curve shows the cross correlation function between GCs and stars in the entire Milky Way as selected from \textit{Gaia} DR3.}
    \label{fig:Skyrecent}
\end{figure*}

\subsection{All escapers}

Across fifty iterations of matching COB populations to Galactic GCs, sampling S+COB encounters in Galactic GCs, and propagating the escaped stars through the Galaxy, we explore the size and makeup of the resultant population of escapers in the present day. We begin by exploring the population as a whole, without regard to their kinematics or detectability. In total, we predict that $6330_{-300}^{+340}$ stars have escaped GCs over the last 500 Myr following S+COB interactions. In Fig.~\ref{fig:StairStep_all} we summarize the distributions of stellar mass, \textit{Gaia G}-band apparent magnitude, and Galactocentric distance in this population. The inner and outer black contours enclose 68\% and 95\% of the total population, respectively. The prototypical escaped S+COB star has a mass of $\sim0.1 \; \mathrm{M_\odot}$ and a magnitude of $G\approx26$. Its total velocity in the Galactocentric rest frame is most likely a few hundred $\mathrm{km \ s^{-1}}$ but can be up to $\sim 2000 \; \mathrm{km \ s^{-1}}$. Many of these stars are currently located in the inner Galaxy -- 13\% and 48\% are confined to the inner 3 kpc and 10 kpc, respectively. Others, however, extend out to the Galactic halo and beyond, reaching distances as far as several hundred kiloparsecs. The blue curve in the $v_{\rm GC}$-vs-$d_{\rm GC}$ plot (bottom row, second from right) shows the escape velocity from the Galaxy in the  \texttt{MWPotential2014} potential -- note that only 18\% of escaped stars are unbound to the Galaxy. 

The panel in the upper right corner of Fig.~\ref{fig:StairStep_all} shows the sky distribution of escaped GC stars in Galactic coordinates. Escaped stars can be found all across the sky, and are not necessarily close on the sky to their parent GC. We delve deeper into this in the uppers plot of Fig.~\ref{fig:Skyrecent}, where we show the sky distributions of stars which escaped from GCs in the recent past ($t_{\rm esc} < 50 \;\mathrm{Myr}$) and split them up by ejection velocity. Escaped stars ejected recently at modest velocities ($v_{\rm esc} < 100 \; \mathrm{km \ s^{-1}}$) remain relatively close to their parent GC on the sky, but the spatial correlation between clusters and their escaping stars begins to break down for larger ejection velocities ($100 \; \mathrm{km \ s^{-1}} < v_{\rm esc} < 500 \; \mathrm{km \ s^{-1}}$) and appears completely absent for the largest ejection velocities. We show this more quantitatively in the lower left plot, where we show the distribution of angular separations between recently escaped S+COB stars and their host GCs using the same binning in ejection velocity. While 77\% of $v_{\rm esc} < 100 \; \mathrm{km \ s^{-1}}$ stars are within 10 degrees of their host GC, this strong clustering drops off steeply with ejection velocity, as only 29\% of $100 \; \mathrm{km \ s^{-1}} < v_{\rm esc} < 500 \; \mathrm{km \ s^{-1}}$ stars and 11\% of $v_{\rm esc} > 500 \; \mathrm{km \ s^{-1}}$ stars are within 10 degrees of their host cluster.

In the lower right plot of Fig.~\ref{fig:Skyrecent} we show the two-point angular cross-correlation function between GCs and recently escaped S+COB stars at a variety of ejection velocities, i.e. the excess probability of finding a S+COB-ejected star at a given angular distance from any GC above the probability one would expect if both populations were distributed randomly and uniformly on the sky. We calculate the cross-correlation function by adapting the two-point autocorrelation function approach provided in the \texttt{astroML} package \citep{VanderPlas2012} and we use a \citet{Landy1993} estimator. For comparison, with the black curve we show the two-point cross-correlation function between GCs and stars in the Milky Way as a whole, where the Milky Way star sample is simply a random subset of the entire \textit{Gaia} DR3 astrometric catalogue. Note that this comparison is not particularly robust since in this section we have not yet accounted for selection effects in our mock S+COB population, but it can provide a useful benchmark nonetheless. We see that among recently escaped S+COB stars, only those escaping with the slowest velocities have an excess probability of being found close to their host cluster on the sky -- this excess drops quickly with escape velocity, and stars escaping faster than $500 \; \mathrm{km \ s^{-1}}$ are not any more likely than a random Milky Way star to be found near a GC.



\subsection{Observable escapers}

\begin{figure*}
    \centering

    \includegraphics[width=\columnwidth]{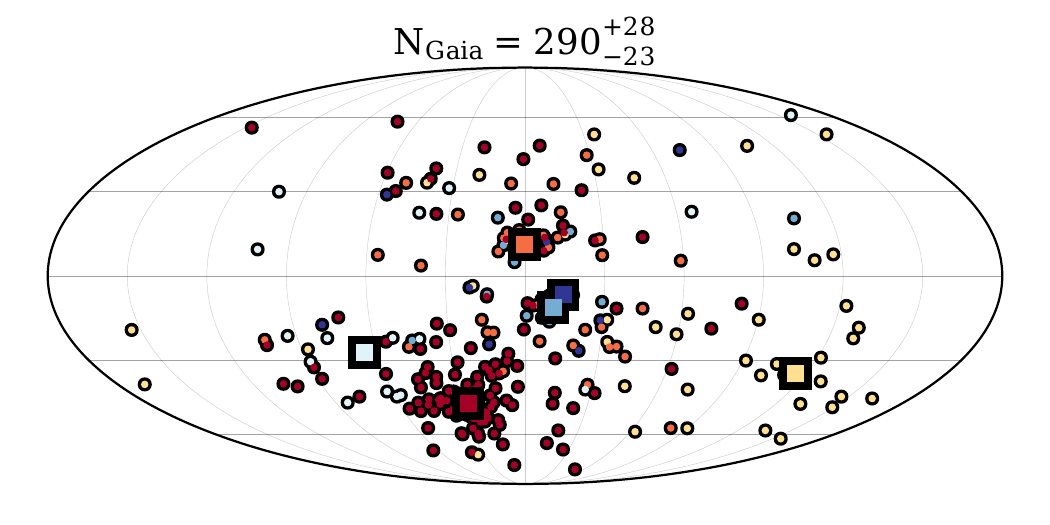}
    \includegraphics[width=\columnwidth]{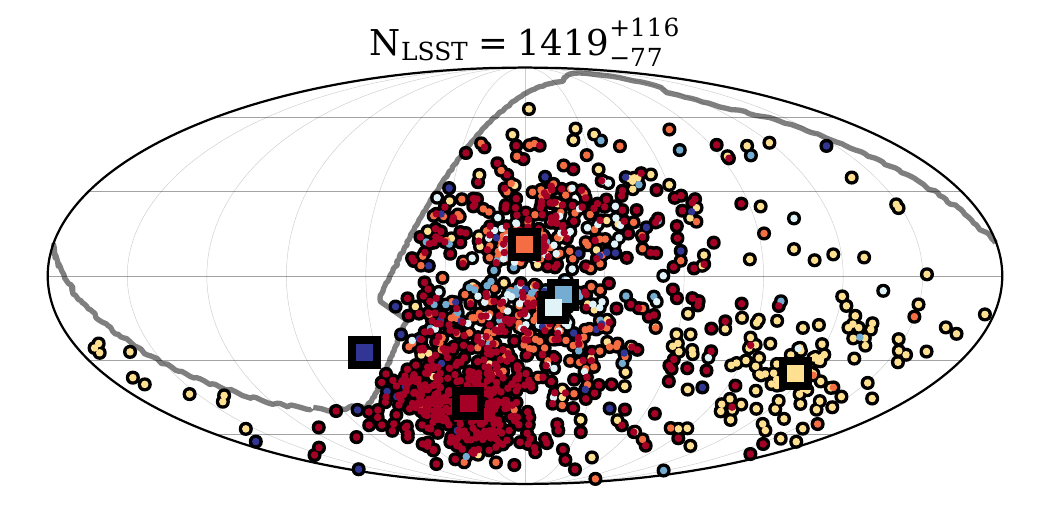}

    \includegraphics[width=\columnwidth]{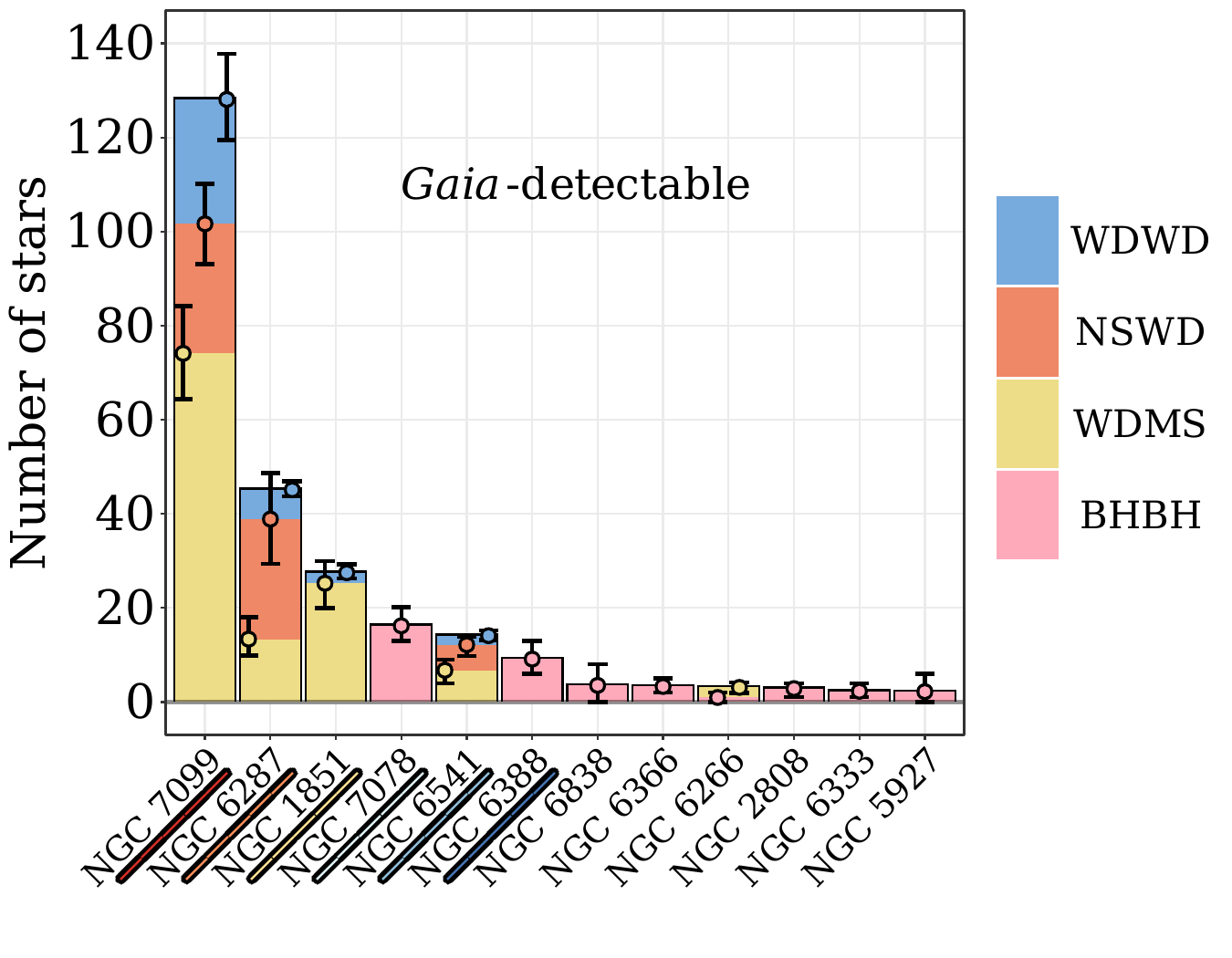}
    \includegraphics[width=\columnwidth]{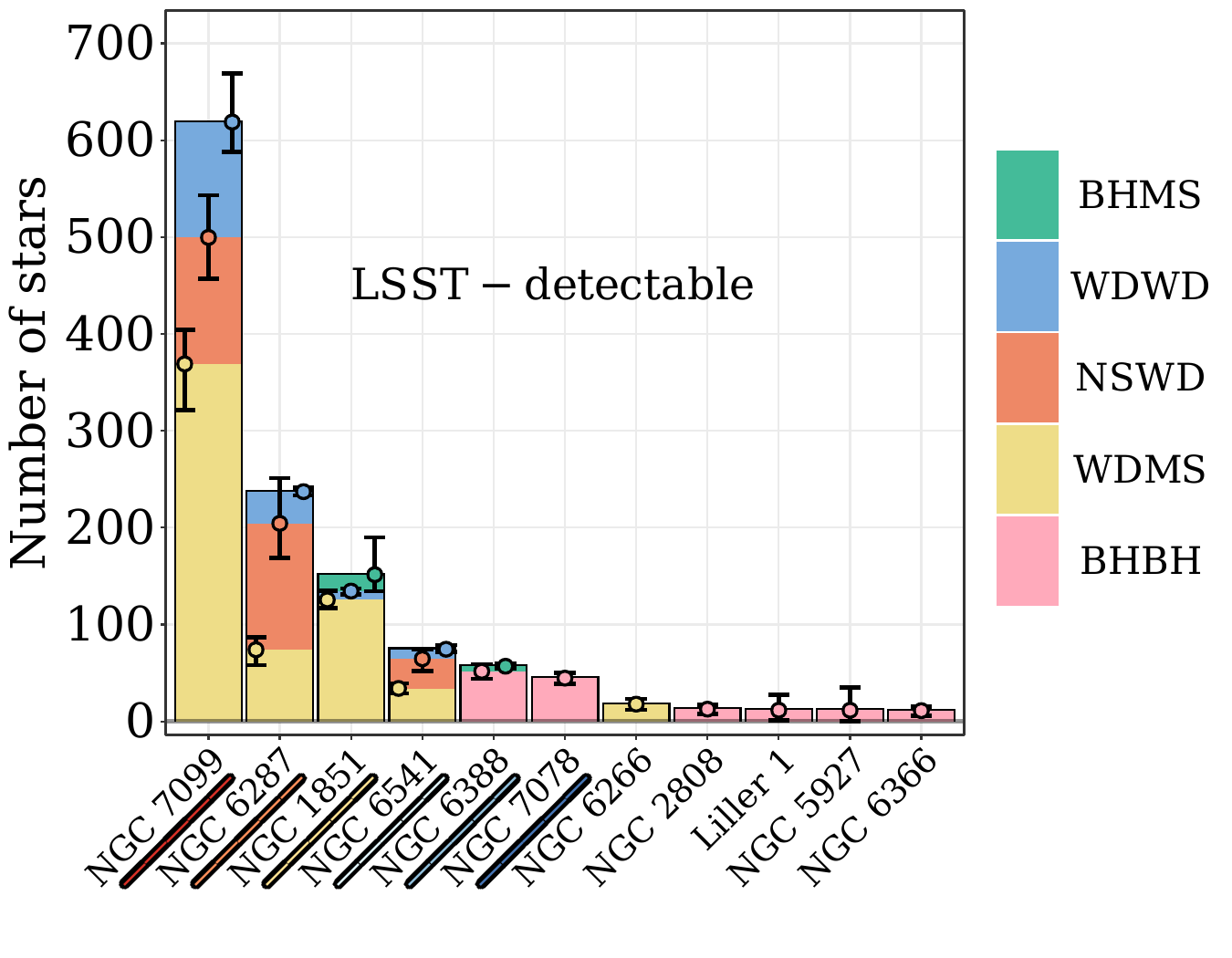}
    
    \caption{\textit{Upper:} Sky distribution of the predicted populations of escaped GC stars detectable by \textit{Gaia} (left) and LSST (right) in Galactic coordinates. Stars share a color with the GC from which they were ejected. Clusters and stars are only shown if they on average eject more than 10 (30) stars detectable by \textit{Gaia} (LSST). The grey curve in the right plot shows the northern boundary of LSST's survey area. \textit{Lower:} The population of detectable stars each Milky Way GC contributes to. The filled portions indicate how much of the total population was ejected following an interaction with each COB type, with the error bars indicating the 1$\sigma$ scatter. The most-ejecting clusters are underlined with the color corresponding to the same cluster in the upper plots. Only clusters which are predicted to eject $>$1 ($>$10) detectable stars are shown in the \textit{Gaia} (LSST) plots.} 
    \label{fig:Skyall}
\end{figure*}

It is clear from Fig.~\ref{fig:StairStep_all} that the majority of escaped stars from S+COB encounters in GCs will be extremely faint -- given the low masses and large Galactocentric distances of the escapers, it is unsurprising that the apparent magnitude distribution peaks $\sim$five magnitudes fainter than the \textit{Gaia} detection limit at $G=20.7$. The red curves and contours in Fig.~\ref{fig:StairStep_all} show the distributions for only the $290_{-23}^{+28}$ stars that we estimate would appear in the \textit{Gaia} DR3 astrometric catalogue. These brighter stars are typically at least $\sim$0.5 $\mathrm{M_\odot}$ in stellar mass, are no farther than $\approx$20 kpc from the Galactic Centre and have velocities no faster than $\approx500 \; \mathrm{km \ s^{-1}}$. The \textit{Gaia}-detectable population tends towards larger stellar masses and tighter distributions of $v_{\rm GC}$ and $d_{\rm GC}$ when compared to the overall population. We stress, however, that these are the stars detectable in the \textit{astrometric} catalogue -- only $2_{-2}^{+0}$ of these would additionally have a radial velocity measurement from \textit{Gaia}. The implications of this are described in greater detail in Sec. \ref{sec:results:observability}. The distributions for the $1419_{-77}^{+116}$ LSST-detectable stars are omitted from this plot for clarity as they occupy similar regions in parameter space as the \textit{Gaia}-detectable population but extend to fainter magnitudes and larger distances. 

The upper left panel of Fig.~\ref{fig:Skyall} shows the sky distribution in Galactic coordinates of these \textit{Gaia}-detectable stars. While escaped stars can still be found quite distant on the sky from their parent cluster, the distribution of \textit{Gaia}-detectable escapers is concentrated near the Galactic meridian ($-60 < l < 60$) and excludes the Galactic midplane where dust attenuation is strongest. 
The upper right panel of Fig.~\ref{fig:Skyall} similarly shows the sky distribution of LSST-detectable ejected stars. Stars can be found spanning LSST's entire sky coverage. Recall that we designate a source as LSST-detectable if it is south of the LSST northern coverage boundary (grey curve) and has a magnitude in the LSST $r$ band of $16<r<24.5$. This upper limit is the faint-end single-visit for a single LSST pointing. Most of the survey area, however, will receive hundreds or thousands of LSST visits over the ten year survey, so the faint-end coadded magnitude limit will be pushed up to three magnitudes deeper \citep{Marshall2017, Ivezic2019}.

In these sky plots we only show clusters from which 10 (30) or more \textit{Gaia-} (LSST-) detectable stars are ejected -- it is clear from this that only a small number of Milky Way GCs are contributing to the greater share of these populations. Of the 149 GCs in our sample, only 18 (51) GCs eject more than one star detectable by \textit{Gaia} (LSST). In the lower panels of Fig.~\ref{fig:Skyall} we further break down how many detectable S+COB stars escape each GC and via which COB type. Trends are quite similar between the two surveys -- $\sim$50\% of both \textit{Gaia}- and LSST-detectable stars are ejected by NGC 7099 alone, and with NGCs 6287, 1851, 7078, 6541 and 6388 joining it, the top six clusters account for 86\% of detectable stars. Half of all S+COB-ejected stars are ejected following interactions with WDMS binaries, with the remainder spread among NSWD, BHBH and WDWD binaries. Notice that NGC 7078 ejects the sixth-most LSST-detectable stars despite the cluster itself lying outside the survey sky coverage.

One might intuitively expect that nearby clusters and clusters in low-density regions of the sky (i.e. far from the Galactic Centre and midplane) would have an advantage when it comes to ejecting detectable stars. These considerations, however, seem to be subdominant to the raw S+COB ejection rate in each cluster. While it is true that the most-ejecting cluster NGC 7099 is in an observationally accessible area of the sky, the clusters which eject the most \textit{Gaia-} or LSST-detectable stars are overall those with the largest S+COB interaction rates -- the six most-ejecting clusters are among the top seven GCs when sorted by mean S+COB interaction rate. The cluster whose most under-represented in \textit{Gaia} DR3 relative to its S+COB ejection rate is Liller 1, which has the sixth-highest interaction rate but ranks 13th among clusters ejecting detectable stars. It being within 1 kpc of the Galactic Centre and strongly affected by dust extinction \citep[see][]{Saracino2015} is a possible explanation for this. 
The distances of the six most-ejecting clusters range from 7.61 kpc (NGC 6541) to 11.95 kpc (NGC 1851), which more or less spans the second tercile of the heliocentric distance distribution of our GC sample. Distance is a likely explanation for the contributions of NGC 6838 and NGC 6366, which are the clusters which eject the seventh and eighth-most \textit{Gaia}-detectable stars despite ranking a humble 23rd and 29th in mean S+COB ejection rate, respectively. At heliocentric distances of 4.0 kpc and 3.44 kpc, these two clusters are the fourth and fifth closest GCs in our sample.

The contributions of each cluster to the LSST-detectable population track more closely with the total S+COB ejection rate. One notable outlier, however, is NGC 7078, which has the second-largest mean interaction rate but ejects only the sixth-most LSST-detectable stars. It is among the most distant of our most-ejecting clusters at 10.71 kpc, but it has the advantage of being in a low-density region of the sky. One possible explanation is that the stars ejected by NGC 7078's BHBH population are travelling so fast that they leave the LSST horizon after a short time -- more on this in the next Section.

\begin{figure*}
    \centering
\includegraphics[width=\columnwidth]{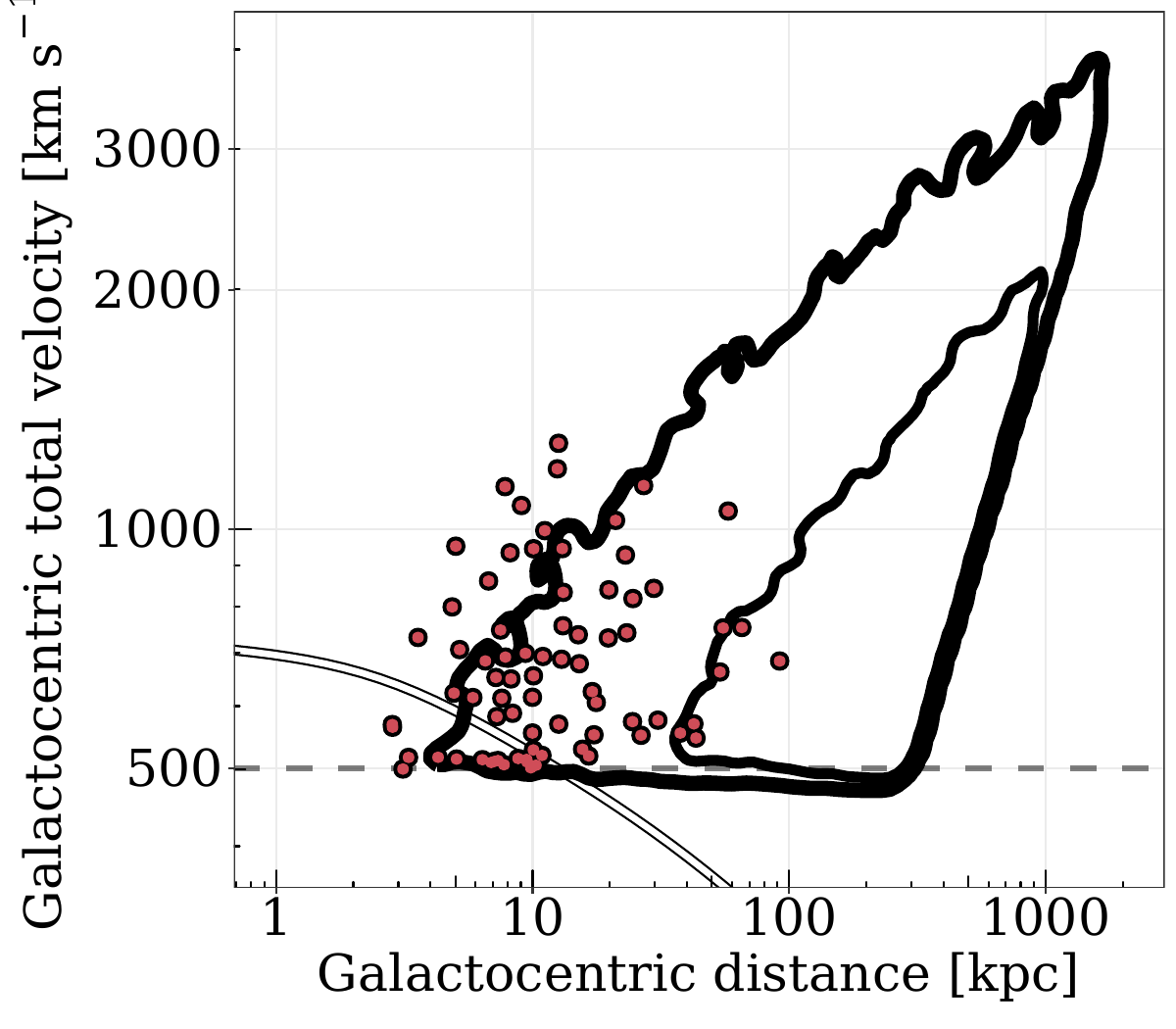}
\includegraphics[width=\columnwidth]{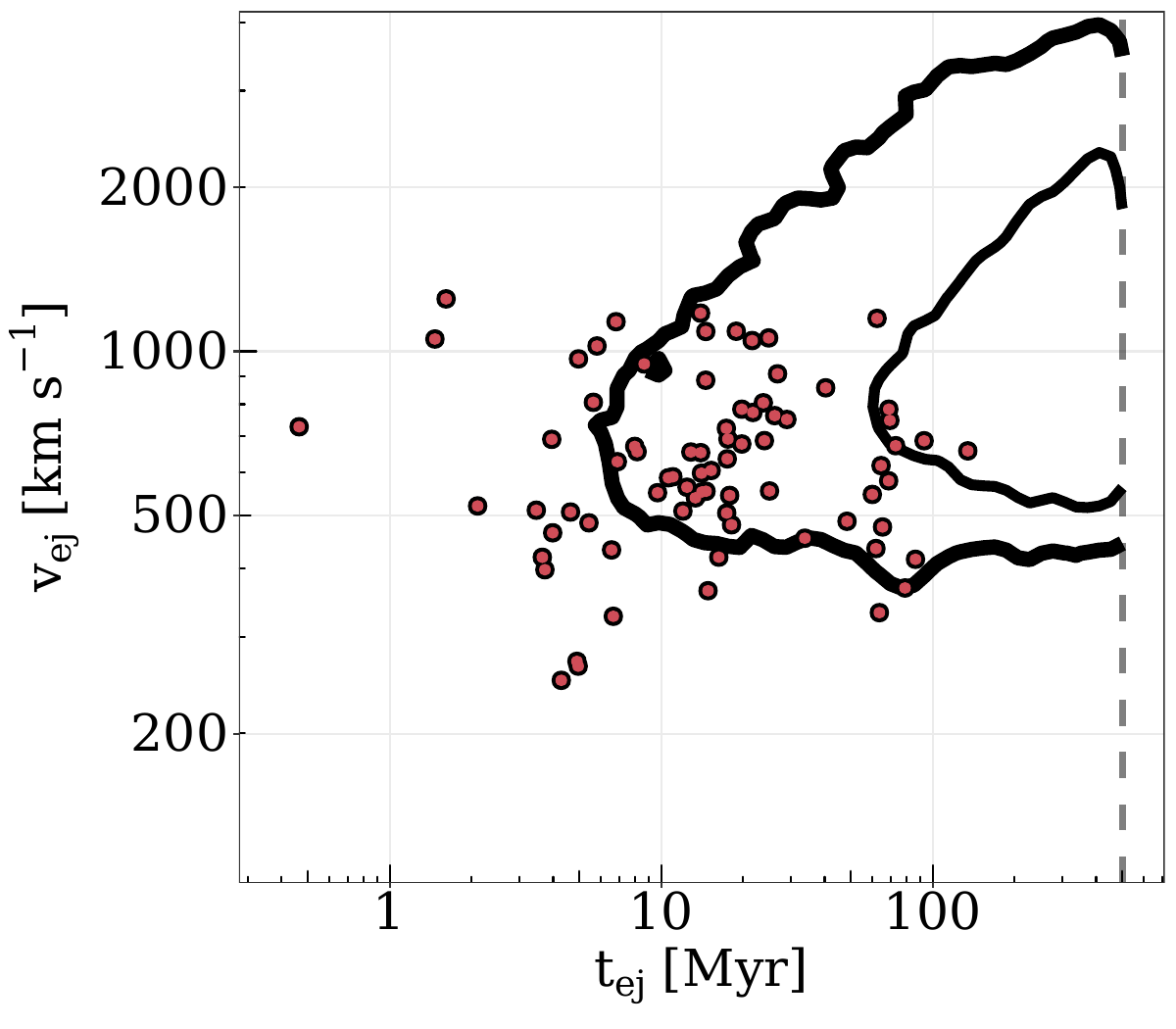}

    \caption{\textit{Left:} The distribution of Galactocentric distances and velocities for escapers with present-day velocities faster than $500 \; \mathrm{km \ s^{-1}}$. Black inner and outer contours enclose 68\% and 95\% of the distribution. Red points show how the $2\pm1$ \textit{Gaia}-detectable stars faster than $500 \; \mathrm{km \ s^{-1}}$ populate this space, stacked over 40 repeated iterations. The white curve shows the escape velocity curve from the \texttt{MWPotential2014} potential. \textit{Right:} The distribution of ejection velocity in the cluster-centric frame and time since ejection for the same population of fast escapers.} 
    \label{fig:summaryfast}
\end{figure*}

\section{Contributions of S+COB interactions to the Galactic hypervelocity star population}
\label{sec:results:fast}

While a large number of stars which escape from GCs following S+COB interactions are expected to be present in the Galaxy and its outskirts, in practice it will be extremely difficult to identify these stars as being associated with GCs, even among stars bright enough to be detectable by \textit{Gaia} or LSST. Without detailed elemental abundances, prospects are not promising for differentiating a GC-escaped star from a typical star in the Galactic halo or bulge using spectral type alone. For this reason, in this section we focus on stars currently travelling faster than $500 \; \mathrm{km \ s^{-1}}$ in the Galactocentric rest frame. While obtaining precise kinematic measurements for these stars will still be quite difficult at least in the near future (more on this in the following Section), given precise measurements these stars would be $>3\sigma$ velocity outliers in both the stellar halo \citep{Bond2010, Kafle2014} and bulge \citep{Portail2015, Portail2017} and would therefore stand out as conspicuous candidates for further study.

Our simulations indicate that over the last 500 Myr, S+COB interactions in Milky Way GCs have ejected $839_{-67}^{+70}$ escaped stars whose present-day velocities in the Galactocentric rest frame exceed $500 \; \mathrm{km \ s^{-1}}$. In the left panel of Fig.~\ref{fig:summaryfast} we show the Galactocentric distance and total velocity distributions of these stars. We show as well how the $1_{-1}^{+2}$ fast stars detectable by \textit{Gaia} populate this space, stacked over all iterations. While the \textit{Gaia}-detectable fast escapers are typically confined to the inner few tens of kpc of the Galaxy and have velocities less than $1000 \; \mathrm{km \ s^{-1}}$, the overall fast population can reach velocities of $>2000 \; \mathrm{km \ s^{-1}}$ and distances of up to a megaparsec. Note that while we have selected the fastest escapers, a minority of this population (0.9\% of the overall and 26\% of \textit{Gaia}-detectable escapers) are still bound to the Galaxy and therefore not genuine  `hypervelocity stars' by most interpretations of the term. A cut at $v=700 \; \mathrm{km \ s^{-1}}$ is a stricter but less ambiguous cut for selecting stars most likely to be unbound from the Galaxy, see \citet{Marchetti2022}. Of our $839_{-67}^{+70}$ fast stars, $567_{-60}^{+38}$ stars satisfy this criterion.

In the right panel of Fig.~\ref{fig:summaryfast} we show the distribution of cluster-frame ejection velocities $v_{\rm ej}$ and the time $t_{\rm ej}$ ago of the ejection, for all stars which are faster than $500 \; \mathrm{km \ s^{-1}}$ today and for the subset of this population detectable by \textit{Gaia}. Detectable fast escapers are outliers in this space -- they have been ejected less than $100 \; \mathrm{Myr}$ ago at a typical ejection velocity of $\sim 500 \; \mathrm{km \ s^{-1}}$. However, stars ejected with velocities as low as $\sim250 \; \mathrm{km \ s^{-1}}$ can still reach Galactocentric velocities greater than $500 \; \mathrm{km \ s^{-1}}$ today in cases where their ejection direction is aligned with the GC's motion and/or they are accelerated by travelling deeper into the Milky Way potential well. The $13_{-4}^{+3}$ fast stars detectable by LSST populate a similar region of parameter space as the \textit{Gaia}-detectable stars, extending to slightly larger distances and ejection times.

\begin{figure*}
    \centering

    \includegraphics[width=1.1\columnwidth]{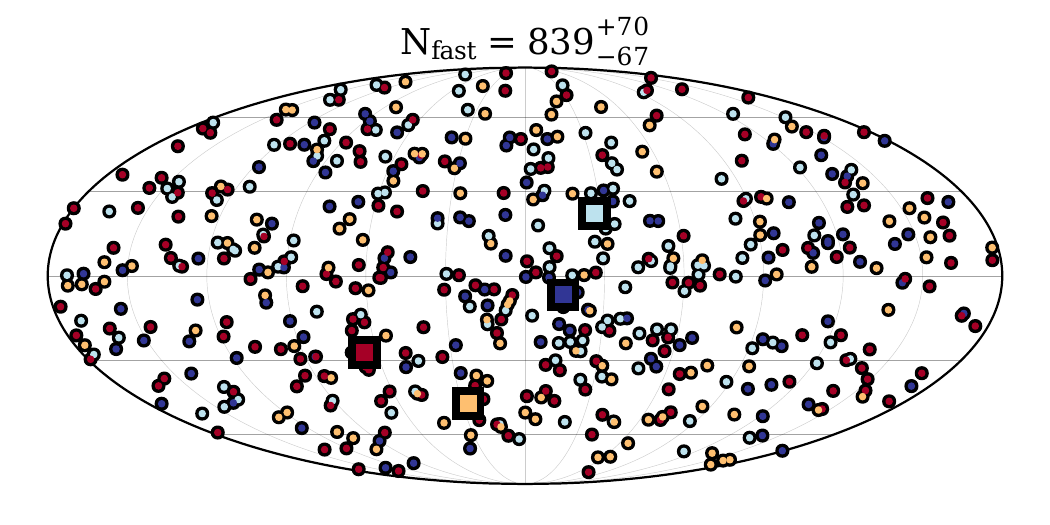}
    
    \includegraphics[width=\columnwidth]{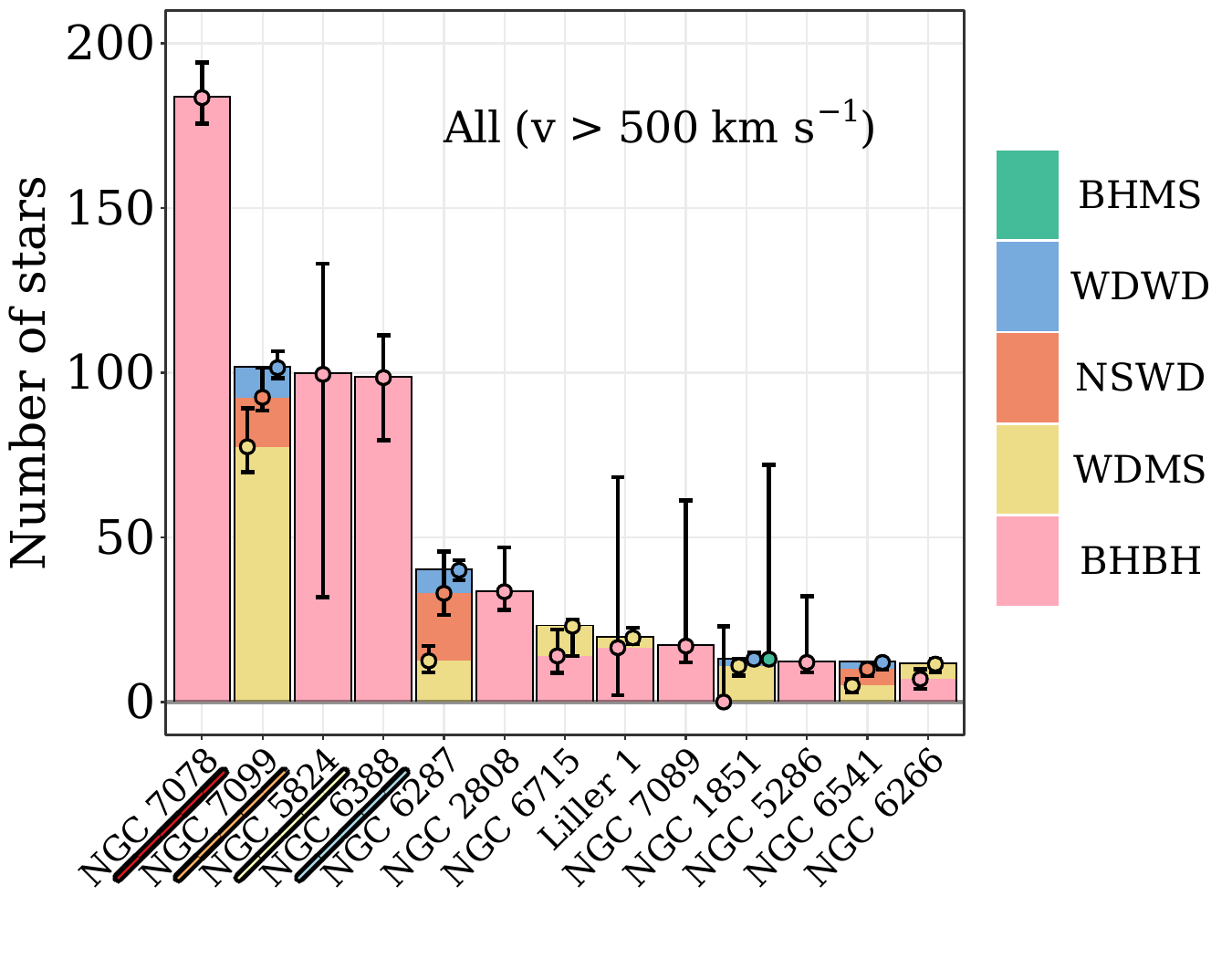}
    \includegraphics[width=0.78\columnwidth]{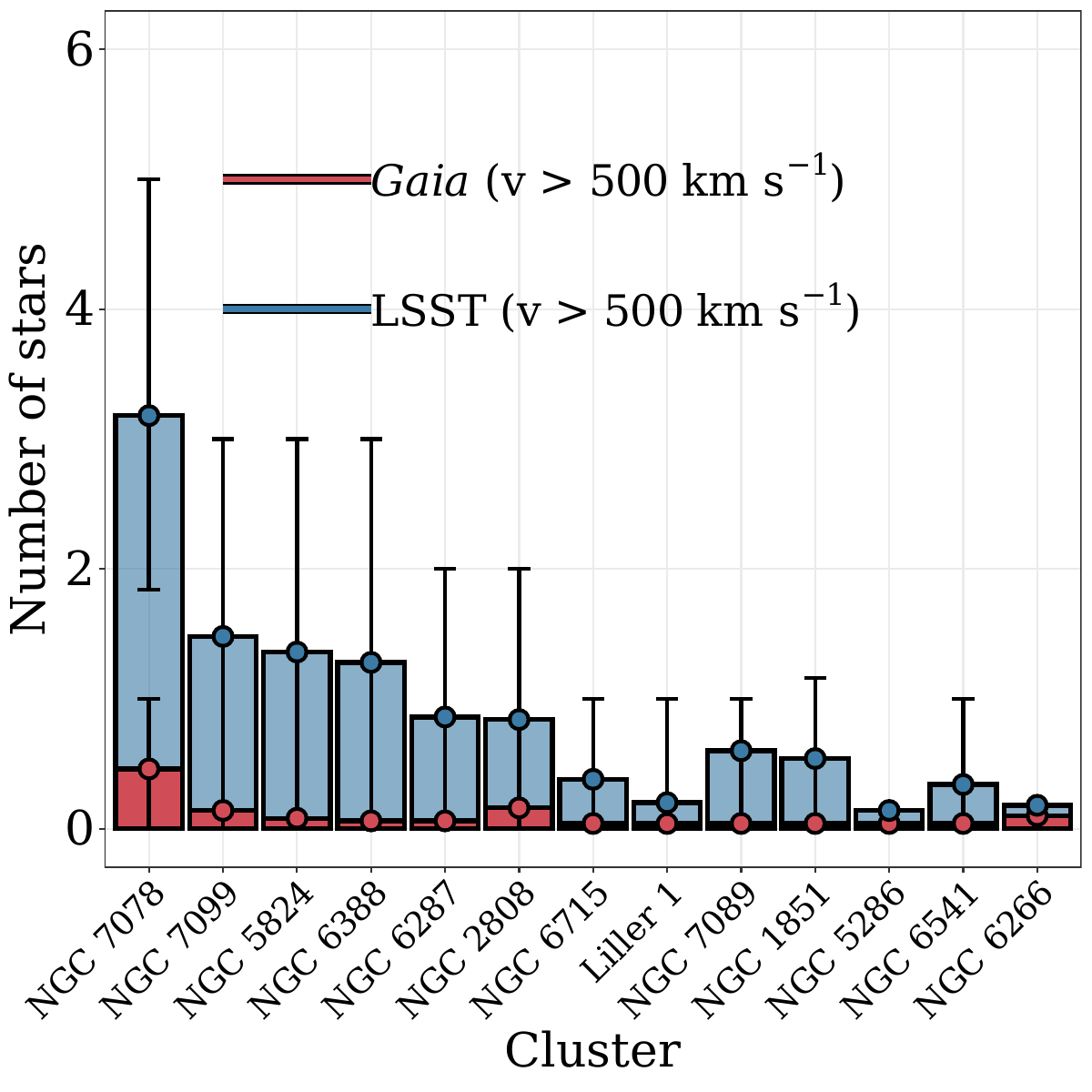}
    
    \caption{\textit{Top:} Sky distribution of fast ($v> 500 \; \mathrm{km \ s^{-1}}$) S+COB-ejected stars in Galactic coordinates. Clusters are only shown if they eject $\geq50$ fast stars on average. \textit{Lower left:} The population of fast stars each Milky Way GC contributes. The filled portions indicate the contribution of each COB type per cluster, with the error bars indicating the 1$\sigma$ scatter over 50 iterations. The most-ejecting clusters are underlined with the color corresponding to the same cluster in the upper plot. Only clustesr which are predicted to eject $\geq$10 stars on average are shown. \textit{Lower right:} The population of \textit{Gaia}- or LSST-detectable fast stars each Milky Way GC contributes.}
    \label{fig:clusternumsfast}
\end{figure*}

\begin{figure}
    \centering
    \label{fig:clusternumscoef}
\end{figure}

\begin{figure*}
    \centering
    \label{fig:clusternumsmodel}
\end{figure*}

\begin{table*} 
\begin{tabular}{|c|cc|cc|cc|cc|} 
    \hline
     cluster & \multicolumn{2}{c}{$Gaia$ DR3} &  \multicolumn{2}{|c|}{LSST} &  \multicolumn{2}{c}{Fast ($v_{\rm GC}\geq500 \ \mathrm{km \ s^{-1}}$)} & \multicolumn{2}{|c|}{total rate $[\mathrm{Myr^{-1}}]$} \\
      & \multicolumn{2}{c}{$N_{\rm total} = 290_{-23}^{+28}$} &  \multicolumn{2}{|c|}{$N_{\rm total} = 1419_{-77}^{+116}$} &  \multicolumn{2}{c}{$N_{\rm total} = 839_{-67}^{+70}$} & \multicolumn{2}{|c|}{} \\\cline{2-9}
    & $N$ & rank & $N$ & rank & $N$ & rank & $\langle \Gamma_{\rm COB}\rangle$ & rank \\
     \hline\hline
     NGC 7099 & $133_{-16}^{+16}$ & 1 & $662_{-95}^{+30}$ & 1 & $111_{-15}^{+16}$& 2 & 2.20 & 1\\
     NGC 6287 & $44_{-6}^{+11}$& 2 & $230_{-22}^{+37}$& 2 & $42_{-6}^{+12}$ & 5 & 0.88& 4\\
     NGC 1851 & $30_{-5}^{+9}$ & 3 & $134_{-13}^{+51}$ & 3 & $15_{-4}^{+78}$ & 10 & 0.99& 3\\
     NGC 7078 & $16_{-3}^{+4}$& 4 & $52_{-10}^{+6}$  & 6 & $184_{-8}^{+10}$ & 1 & 1.02 & 2\\
     NGC 6541 & $15_{-5}^{+4}$ & 5 & $79_{-18}^{+11}$ & 4 & $13_{-3}^{+4}$ & 12 & 0.30 & 7\\
     NGC 6388 & $10_{-4}^{+5}$& 6 & $55_{-7}^{+9}$ & 5 & $99_{-19}^{+14}$ & 4 & 0.41 & 5\\
     NGC 5824 & $0.6_{-0.4}^{+0.6}$& 33 & $9_{-6}^{+5}$ & 12 & $100_{-68}^{+34}$ & 3 & 0.17 & 11\\
     \hline
\end{tabular}
\caption{The number of expected \textit{Gaia} DR3-detectable, LSST-detectable and $v>500 \; \mathrm{km \ s^{-1}}$ stars expected for the most-ejecting GCs, along with their rankings among all clusters and their total S+COB ejection rates.}
\label{tab:counts}
\end{table*}


In the the top panel of Fig. \ref{fig:clusternumsfast} we show how the $v \geq 500 \, \mathrm{km \ s^{-1}}$ S+COB-ejected stars are distributed across the sky. As suggested by the spatial distribution of the fastest-ejected stars in Fig. \ref{fig:Skyrecent}, we see that these fast stars can be found all across the sky and not necessarily close at all to their host cluster. In the lower left panel of Fig.~\ref{fig:clusternumsfast}, we show how the population of $v \geq 500 \, \mathrm{km \ s^{-1}}$ S+COB-ejected stars are distributed among the Galactic GCs and the different COB types. NGC 7078 ejects 22\% of all fast stars, and together with NGCs 7099, 5824 and 6388, the top four clusters eject 57\% of all fast stars. These four clusters are among the same clusters which contribute most to the overall \textit{Gaia-} and LSST-detectable population (Fig.~\ref{fig:Skyall}) with the exception of NGC 5824, which ejects the third-most fast stars on average (albeit with large scatter and more or less tied with NGC 7099 and NGC 6388) despite having only the 11th largest overall S+COB ejection rate. It is clear from these plots that S+BHBH interactions are associated with fast ejections. 73\% of escaped fast stars are ejected following interactions with BHBH binaries, whose large binding energies lead to a more extended high-velocity tail of the single star escape velocity distribution (see Fig.~\ref{fig:vej}). The remainder of fast ejections primarily follow interactions with WDMS binaries in NGC 7099. The large error bars shown by some clusters, particularly BHBH ejections in NGC 5824, Liller 1 and NGC 7089, as well as BHMS ejections in NGC 1851, show that predictions for these clusters are quite sensitive to which \texttt{CMC} model output is matched to them -- we comment on this more in Sec. \ref{sec:discussion}.

In the right panel of Fig.~\ref{fig:clusternumsfast}, we show the number of \textit{Gaia-} and LSST-detectable fast escaped stars ejected from each cluster, preserving the ranking from the left panel. While we expect on average $1_{-1}^{+2}$ \textit{Gaia}-detectable fast stars, no single GC is expected to consistently eject $\gtrsim$1 detectable star. NGC 7078 seems the most promising with a $\sim$50\% chance of ejecting a fast and \textit{Gaia}-detectable star -- the other most-ejecting clusters \textit{can} eject fast \textit{Gaia}-detectable escaped stars, but this occurs at a few times in 50 iterations of each cluster. All of the top four ejecting clusters are expected to eject between one and three LSST-detectable fast stars, but with significant uncertainties. 

In Table \ref{tab:counts} we summarize the results of Secs. \ref{sec:results:slow} and \ref{sec:results:fast}. For each of the most important GCs we have mentioned in these sections, i.e. the GCs which eject the most detectable stars in \textit{Gaia}/LSST as well as NGC 5824 which contributes significantly to the fast S+COB population, we recap the number of stars each cluster contributes to each population and its rank among all GCs. For reference, we also show the mean S+COB ejection rate we calculate for each GC.       

\section{Can we find S+COB stars in Gaia?} \label{sec:results:observability}

\begin{figure*}
    \centering
\includegraphics[width=2\columnwidth]{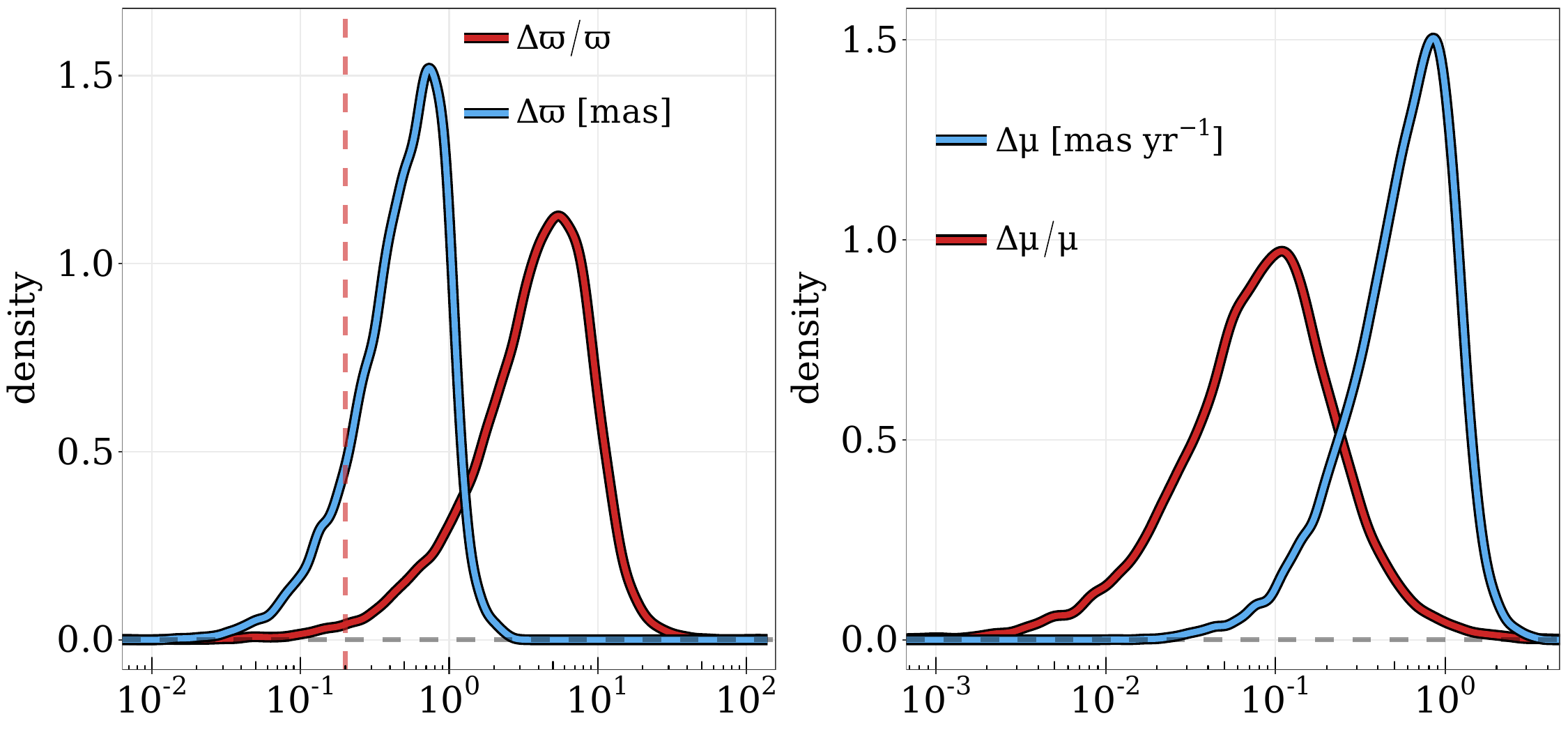}
    \caption{Distributions of the relative (red) and absolute (blue) parallax (left) and proper motion (right) errors among \textit{Gaia}-detectable S+COB-ejected stars. The vertical dashed line marks a relative parallax error of 20\%, above which distance estimation using parallax alone becomes non-trivial.}
    \label{fig:GaiaErrors}
\end{figure*}

\begin{figure*}
    \centering
    \includegraphics[width=\columnwidth]{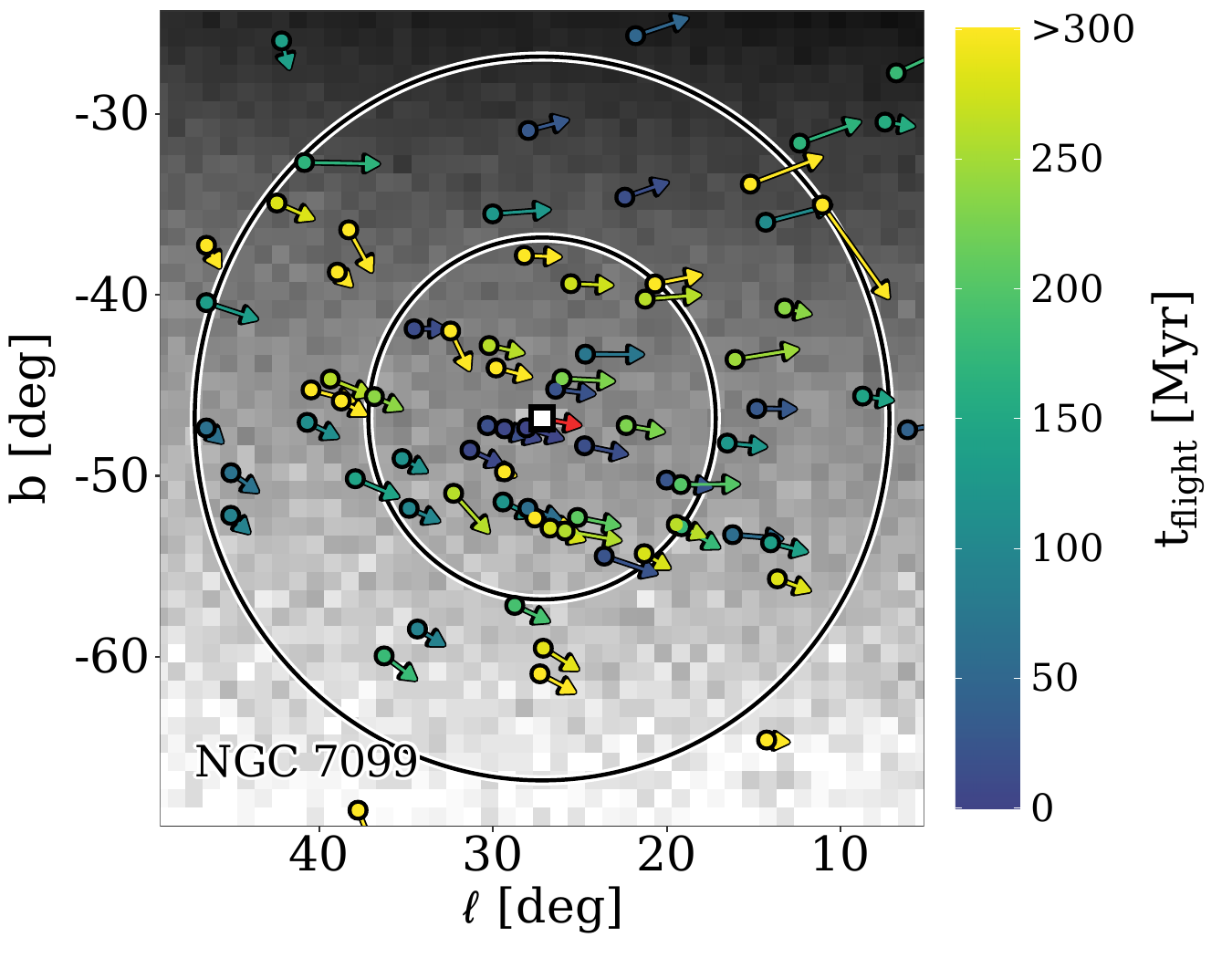}
    \includegraphics[width=0.9\columnwidth]{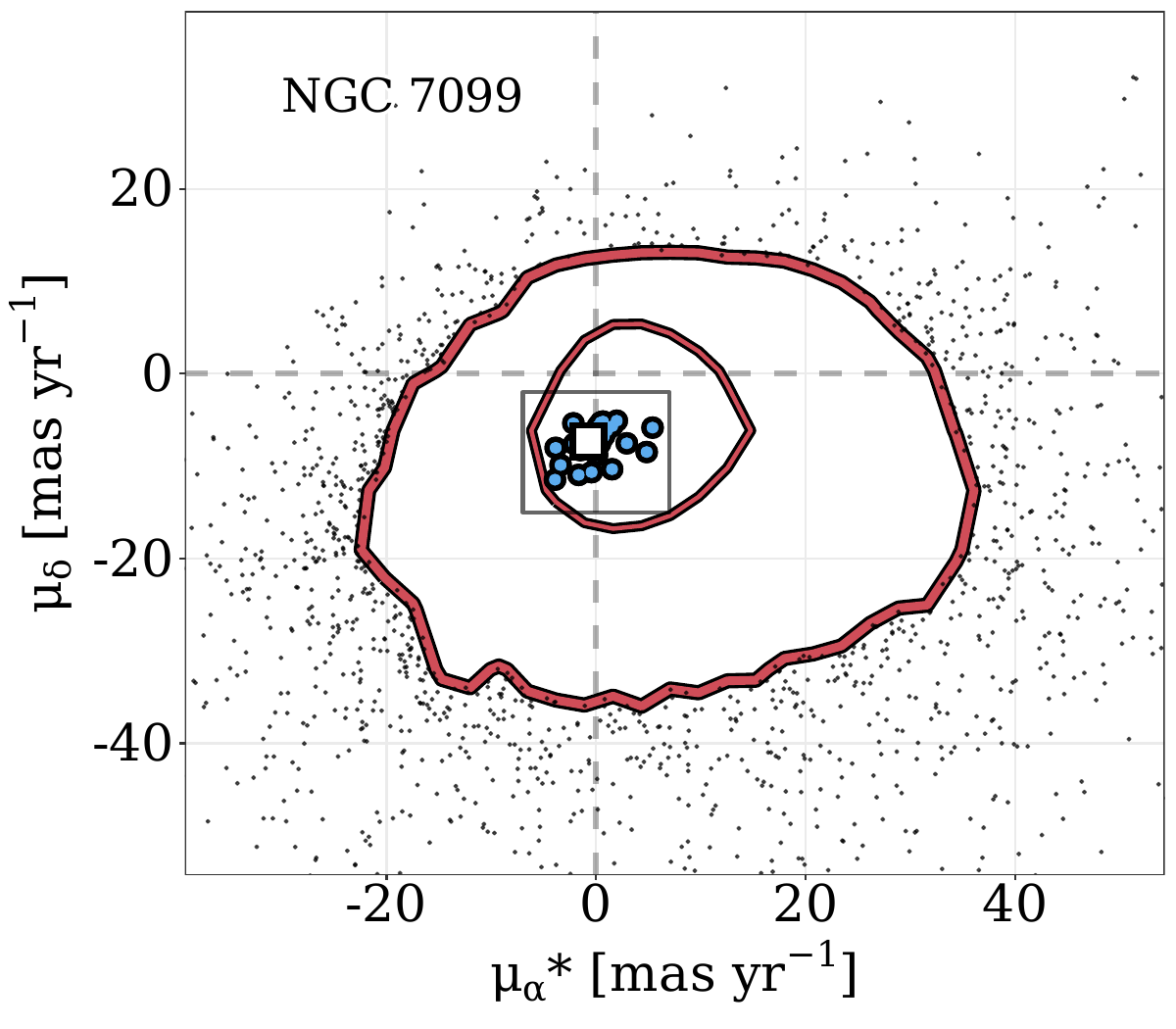}
    \caption{\textit{Left:} Equirectangular zoom-in on the $40^\circ$x$40^\circ$ field centred on NGC 7099. Escaped stars are colored by their time since ejection. The vectors indicate the magnitude and direction of each star's proper motion on the plane of the sky. The white square and red vector show the location and proper motion of NGC 7099. In greyscale in the background is the log-density of sources near NGC 7099 in \textit{Gaia} DR3. \textit{Right}: The blue points show the predicted proper motion components of S+COB stars ejected by NGC 7099 within $10^\circ$ of it. The grey box bounds the distribution of these points (see text for details). The white square shows the proper motion of NGC 7099 itself. The contours show the 68\% and 95\%  density contours for stars within 10$^\circ$ of NGC 7099 in \textit{Gaia} DR3. The proper motions of stars outside the outermost contour are plotted with individual points.} 
    \label{fig:70786397}
\end{figure*}

We  established in the previous Sections that our modelling suggests that $290_{-23}^{+28}$ S+COB-ejected escaped stars are currently in the \textit{Gaia} DR3 catalogue, and $1_{-1}^{+2}$ of these have velocities above $500 \; \mathrm{km \ s^{-1}}$. Whether or not it is feasible to  \textit{identify} individual stars from among the $\sim$two billion sources in the catalogue as being S+COB-ejected stars from particular GC is another matter. In this Section we discuss the observational realities of trying to actually identify these stars. 

If their positions and velocities are known perfectly, propagating these stars backwards in time could allow one to associate them with their parent clusters, at least for fast and/or recently escaped stars which are less sensitive to uncertainties in the Galactic potential. Recall, however, that only $2_{-2}^{+0}$ of these stars would have a \textit{Gaia}-measured radial velocity. Significant astrometric uncertainties would further hamstring our knowledge of the kinematics of these stars. In Fig.~\ref{fig:GaiaErrors} we show the distributions of the parallax errors (left) and proper motion errors (right) for \textit{Gaia}-detectable stars as estimated by \texttt{GaiaUnlimited}.  While proper motion errors are, fairly small, peaking at $\sim$ 1 $\mathrm{mas \ yr^{-1}}$ or $\sim$10\%, parallax errors for these stars would be on the order of several hundred $\mu$as, which at these stars' distances is an error several times greater than the measured parallax itself. Estimating distances via inverting parallaxes becomes non-trivial for relative uncertainties above $\sim$20\% \citep{BailerJones2015}. Only 1\% of the \textit{Gaia}-detectable population ($\sim$3 stars) achieves this, that is to say, the heliocentric distance to the vast majority of \textit{Gaia}-detectable S+COB-ejected escaped stars will be quite poorly constrained. 

While in this work we have focused primarily on \textit{Gaia} DR3 and LSST, it is worth a brief diversion to explore the population of S+COB stars in \textit{Gaia} Data Release 4 (expected $\sim$2026). The \textit{Gaia} DR4 astrometric catalogue will not include many more S+COB stars than DR3 since both extend down to \textit{Gaia's} faint-end magnitude limit at $G\approx20.7$. However, a longer time baseline will lead to parallax precision improvements by a factor of 1.3 relative to DR3\footnote{see \url{www.cosmos.esa.int/web/gaia/science-performance}}. Even so, this improvement will still only leave us with $\sim$5 stars in the Data Release with parallax errors below 20\%. Improvements in the radial velocity validation pipeline, though, mean that radial velocity measurements should be available for all cool $T_{\rm eff}\leq6900$ K stars down to the $G_{\rm RVS}\approx16.2$ limiting magnitude of the \textit{Gaia} RVS spectrometer \citep{Katz2019} and for all hotter stars down to $G_{\rm RVS}=14$. We determine which of our mock S+COB-ejected stars will appear in the \textit{Gaia} DR4 radial velocity catalogue by computing each star's mock $G_{\rm RVS}$ magnitude from its $I_c$, $V$, and $Gaia \ G$ magnitudes using the fitting functions of \citet{Jordi2010} (Table 3). We find that $28_{-3}^{+6}$ stars will satisfy the temperature and $G_{\rm RVS}$ criteria to appear in the \textit{Gaia} DR4 radial velocity catalogue. This is significantly more than the maximum of two in the DR3 radial velocity, but still not a particularly large sample.

In the remainder of this section we focus more on NGC 7099, the Milky Way GC which contributes the plurality (48\%, or $133_{-16}^{+16}$) of \textit{Gaia} DR3-detectable S+COB-ejected stars and ejects the second most fast stars ($111_{-15}^{+16}$). In the left panel of Fig.~\ref{fig:70786397} we show the 40$^\circ$ x 40$^\circ$ area centred on NGC 7099. The points show \textit{Gaia}-detectable S+COB stars within this area. Within $10^\circ$ of NGC 7099 we predict $30_{-5}^{+4}$ \textit{Gaia}-detectable escaped S+COB stars. In greyscale in the background is the logarithmic count density of sources in \textit{Gaia} DR3; the lightest and darkest bins correspond to source densities of $\sim1100 \; \mathrm{deg^{-2}}$ and $\sim 30,500 \; \mathrm{deg^{-2}}$ in the field of view around NGC 7099. In total there are nearly two million \textit{Gaia} DR3 sources within 10 deg of NGC 7099 but outside its tidal radius.

The vector on each star indicates the direction and magnitude of its proper motion. Note that despite escaping from NGC 7099, many stars were ejected so long ago that their present-day proper motion vectors may not point directly away from NGC 7099. In the right panel of Fig.~\ref{fig:70786397} we show how these detectable stars populate proper motion space, and we show as well the 68\% and 95\% contours for the proper motions of all sources in \textit{Gaia} DR3 located within $10^\circ$ of NGC 7099.  Despite the fact that most of the stars were ejected at considerable velocities, S+COB stars escaped from NGC 7099 would not be large proper motion outliers relative to nearby stars on the sky. S+COB-ejected  stars near NGC 7099 have proper motions confined to the range $-15 \; \mathrm{mas \ yr^{-1}} \lesssim \mu_\delta \lesssim -2 \; \mathrm{mas \ yr^{-1}} $ and $-7 \; \mathrm{mas \ yr^{-1}} \lesssim \mu_\alpha^* \lesssim +7 \;\mathrm{mas \ yr^{-1}}$ which we show using the grey box in the Figure. Querying \textit{Gaia} DR3 for stars within 10 deg of NGC 7099 but outside its tidal radius with proper motions only within these bounds yields 990,000 stars. Some foreground stars can be removed by recognizing that S+COB stars should have large relative parallax errors given their rather large heliocentric distances (Fig.~\ref{fig:GaiaErrors}). Removing stars with relative parallax errors below 20\% leaves 950,000 stars. Finally, we find that S+COB stars ejected from NGC 7099 occupy a \textit{Gaia} $G-G_{\rm RP}$ color range of $[0.3,0.9]$ -- applying this cut to our \textit{Gaia} query as well reduces the sample down to 720,000 stars. 



While more sophisticated mining of \textit{Gaia} DR3 could certainly reduce this sample further, this rough approach serves as an exercise to show that picking out genuine S+COB stars from \textit{Gaia} DR3 (photo)astrometric data alone may be quite difficult in practice. Additional observational evidence is likely required to conclusively determine the origin of S+COB-ejected star candidates. Tracing the origins of Milky Way field stars is often done through chemical tagging: a method that compares the chemical abundances of an observed star with the overall chemical makeup of a specific environment. Chemical tagging has been successful in associating halo stars with individual GCs \citep[e.g.][]{Martell2016, Koch2019, Chun2020, Grondin2023, Kane2025}, largely because the Milky Way GC population generally has larger abundance spreads compared to the open star cluster population \citep[making chemical tagging for these clusters more uncertain][]{Casamiquela2021}. As shown in Figure 6, we find that stars recently escaped from GCs are low mass ($m\sim 0.1-1 M_\odot$) and faint ($G\sim 18-35$ mag). Hence, identifying HVSs in current large scale chemical abundance surveys will be challenging, given the observational capabilities of e.g. APOGEE (primarily observes red giants) and LAMOST (limiting magnitude of $\sim 17$ mag). While high-dimensional chemical tagging might be difficult for associating HVSs with individual GCs, targeted spectroscopy of identified HVSs could be helpful, where a metallicity measurement alone could offer some distinction between different GCs \citep{Harris2010}.

\section{Discussion}
\label{sec:discussion}

\subsection{Model assumptions} \label{sec:discussion:model}

The results in this work depend on a number of assumptions and measurements, many of which are subject to uncertainties. The estimated structural parameters of Milky Way GCs are particularly important, since they dictate the \texttt{CMC} model matched to each GC and factor directly into the S+COB interaction rate (Eq.~\ref{eq:Gamma} and Fig.~\ref{fig:StairStep_cluster}). These structural parameters are taken from the catalog of \citet{Baumgardt2018}, who fit detailed N-body models to velocity dispersion profiles derived from VLT/Keck observations. The relevant parameters (core radius, central density, velocity dispersion) are not published with uncertainties, and can vary significantly with time if new data becomes available and the best-fit model changes. As an extreme example, in the fourth and most recent version of the catalog (updated March 2023), the nearby globular cluster NGC 6397 has an estimated central density of $1.6\times10^{5} \; \mathrm{M_\odot \ pc^{-3}}$, whereas in the prior version of the catalog it was the third most centrally-dense GC in the Galaxy, with a density estimated at $2.5\times10^{6} \; \mathrm{M_\odot \ pc^{-3}}$.


Our modeling also assumes a single present-day stellar mass function (MF) for all clusters. In reality, clusters are known to have a wide range of stellar mass functions \citep{Baumgardt2023} that varies within the cluster itself \citep{Webb2017, Baumgardt2017} due to the combined effects of mass loss \citep{Webb2015} and two-body relaxation \citep{Spitzer1969, Spitzer1987}. Furthermore, direct comparisons between observations and theoretical models even suggest that the initial mass function (IMF) may not be truly universal and could be top-heavy relative to a \cite{Kroupa2001} IMF for some clusters \citep{Cadelano2020,Baumgardt2023}. In the absence of homogeneous and complete stellar mass function estimates, these factors can't be incorporated into a study of the entire Galactic globular cluster population. However, we ce can speculate on how these factors may influence individual clusters.

Switching to a more top-heavy core MF increases the average stellar mass, which reduces the estimated S+COB ejection rate (Eq.~\ref{eq:Gamma}), and decreases the mean ejection velocity of S+COB-ejected stars. Our predictions for the number of fast S+COB stars may represent upper limits, as these two factors would both decrease the number of S+COB stars reaching $v\geq 500 \ \mathrm{km \ s^{-1}}$. However, although fewer stars would be ejected, the heavier (and therefore brighter) stars would be more easily detectable, so the number of \textit{Gaia-} or LSST-detectable stars is less sensitive to the GC core stellar mass function by comparison.

Throughout this work we also assume a fixed binary fraction, which may impact results as well. Binary fraction (which we assume to be $\approx$multiplicity fraction) measurements are available for some but not all GCs in our catalog \citep{Milone2012, Ji2015}, but vary spatially within a cluster and depend on the mass ratio of the binary and the mass of the primary. A flat fraction of 5\% is consistent with a typical Galactic GC. The S+COB ejection rate scales linearly with $(1-f_{\rm mult})$, as more stellar binaries means fewer single stars are available for S+COB interactions. \citet{Milone2012} find that NGC 7099 has a binary fraction of 7\% $\pm$ 3\% within its core and NGC 7078 has a binary fraction of 2\% $\pm$ 1\% within the half-mass radius. The binary fractions for both clusters within the half-mass radius as measured by \citet{Ji2015} vary from 0.5\% to several percent depending on the method.

\subsection{Matching to \texttt{CMC} models} \label{sec:discussion:model}

Our matching of Galactic GCs to pre-existing model outputs (see Sec. \ref{sec:methods:gcs}) also introduces some uncertainty as well, since our predictions for each cluster depend on the COB population of the \texttt{CMC} model matched to it. Recall that we match GCs by first matching to the closest \texttt{CMC} models in log$Z-d_{\rm GC}$ space, and we then select one model timestep at random out of the three closest in log$M-r_{\rm c}/r_{\rm hl}$ space. We test whether our specific GC matching approach impacts our predictions by running two additional suites of 50 iterations, one in which we match to one of the closest \textit{five} \texttt{CMC} models in log$M-r_{\rm c}/r_{\rm hl}$ space instead of the closest three, and one in which we only match to the closest model. These tests show that our predictions for S+COB population are fairly robust against our specific modelling choices. Whether choosing from among the five closest models or choosing only the closest model, the number of \textit{Gaia-} and LSST-detectable S+COB stars changes by less than 10\% and does not change the ordering of the clusters which contribute most to these populations. The number of fast stars ejected is somewhat more sensitive. Choosing from among the five closest models increases the predicted size of the fast S+COB population by 13\% to $938_{-143}^{+416}$, where the large upper uncertainty is due to the fact that the two extra \texttt{CMC} model outputs now available to be matched to NGC 7078 allow it to eject up to $\sim$750 fast S+COB stars. Choosing only the closest model, however, yields a similar predicted fast S+COB population as our default scheme ($871_{-16}^{+13}$), where the main difference is that NGC 1851 is always (rather than in one third of iterations) matched to a \texttt{CMC} output with a BHMS population amenable to fast S+COB ejections.

Recall that in our model matching scheme, Milky Way GCs can be matched to any \texttt{CMC} output timestep between 9 Gyr and 14 Gyr. We have not accounted for cluster age in our matching scheme because a) age estimates are not available for all GCs in our sample, b) even when available, GC age estimates are prone to large uncertainties and systematic biases to photometric uncertainties, calibration errors, assumptions about cluster distance and foreground reddening, and choice of stellar evolution prescription \citep[see][]{Massari2019}. Even so, the ages of many Milky Way GCs can at least be constrained to a narrower range than 9 Gyr to 13 Gyr, and it is worth checking if the \texttt{CMC} outputs matched to our GCs are consistent with these constraints. Of the seven most impactful clusters listed in Table \ref{tab:counts}, NGCs 7099, 6287, 7078, 6541 and 5824 are uncontroversially considered older than $\sim$12 Gyr \citep[see][]{Salaris2002, Santos2004, DeAngeli2005, Meissner2006, Koleva2008, MarinFranch2009, VandenBerg2013}. NGC 6388 is also likely an older cluster \citep[11.5-12 Gyr]{Moretti2009, Massari2023}, and NGC 1851 is typically estimated at closer to 9-10 Gyr old \citep{Salaris2002, Meissner2006}, though precise age estimates for both clusters are hamstrung by their multiple stellar populations and peculiar stellar abundances \citep[see][]{Milone2008, Bellini2013, Tautvaisiene2022, Dondoglio2023, Carretta2023}.

Of the five `old' GCs, only NGC 5824 and NGC 6541 are never matched to a \texttt{CMC} output older than 12 Gyr. For NGC 7099/7078/6287, one/two/three of the three closest \texttt{CMC} models are older than 12 Gyr. To check the effect of age considerations, we run another suite of 50 iterations using just these five clusters, but we match them only to one of the three closest \texttt{CMC} model outputs older than 12 Gyr. The numbers of either detectable or fast S+COB stars ejected remains fairly unchanged with the exception of NGC 7078, which ejects $\sim$80\% more detectable or fast S+COB stars on average, and more notably NGC 6541, which ejects 35x more \textit{Gaia-} or LSST-detectable stars and 50x more fast stars on average. In this prescription, it ejects $978_{-841}^{+51}$ fast stars, twice as much as the other four clusters combined, since two of the three \texttt{CMC} models matched to it contain BHBH populations very amenable to fast ejections.

To sum up, while our overall results are not particularly sensitive to our modelling choices, specific predictions for individual clusters are still subject to uncertainty and should be taken with a grain of salt. In the future, simulations tailored to each GC would provide more robust and reliable estimates of the fast and detectable populations of stars they eject.

\subsection{Previous work}

\citet{Cabrera2023} previously investigated ejecta from interactions between stars and compact object binaries in Galactic GCs. While they also make use of \texttt{CMC} models and match models to observed GCs in a similar way, there are a number of methodological differences. Whereas we compute a fixed S+COB interaction rate for each COB type in each GC over the recent Galactic past, they compute the interaction probability for each COB in each simulation at each timestep. They explicitly compute the interaction using \texttt{Fewbody} \citep{Fregeau2004}, tailored for small-N body systems, choosing a random geometry for the interaction. We instead sample the interaction using \texttt{corespray}, which is faster and thus allows us to sample a high number of interactions spanning a range of parameter space. While they do not account for the detectability of the escapers and do not describe in detail which clusters eject the most stars, their full simulation outputs are fortunately publicly available. To make a more direct comparison, we obtained mock \textit{Gaia} photometry for their outputs following Sec.~\ref{sec:methods:phot} and count the fast and/or \textit{Gaia}-detectable stars as we do with our own simulations. 

\citet{Cabrera2023} find that S+COB-ejected escaped HVSs are rarer, with only $\sim$700 stars with present-day velocities faster than 500 $\mathrm{km \ s^{-1}}$ ejected over the lifetime of the Milky Way, most of these early in its history. Even so, they also find that NGC 7099 ejects the most fast stars of any cluster, and NGC 2808 (the cluster which ejects the eight-most fast stars in our analysis) is also among their most-ejecting clusters. The other clusters most responsible for fast ejections in their outputs are the NGCs 5272, 5904 and 6752, which respectively rank as the 30th, 44th and 62nd most-contributing cluster in our predictions. Only $\sim$130 stars escaped from GCs within the last 500 Myr are \textit{Gaia}-detectable among their simulations, a reduction of 55\% relative to our results. These predictions are similar in magnitude, however, we share no clusters in common among those which contribute most to the detectable S+COB population -- their largest contributors are the NGCs 6535, 4147, 6540, 3201 and 6093. Again, none of these clusters are particularly close nor centrally dense. As mentioned throughout this section, however, we note that predictions for individual clusters are sensitive to assumptions and choices in the modeling approach. 

\subsection{COB escape from clusters}

\begin{figure}
    \centering
    \includegraphics[width=0.95\linewidth]{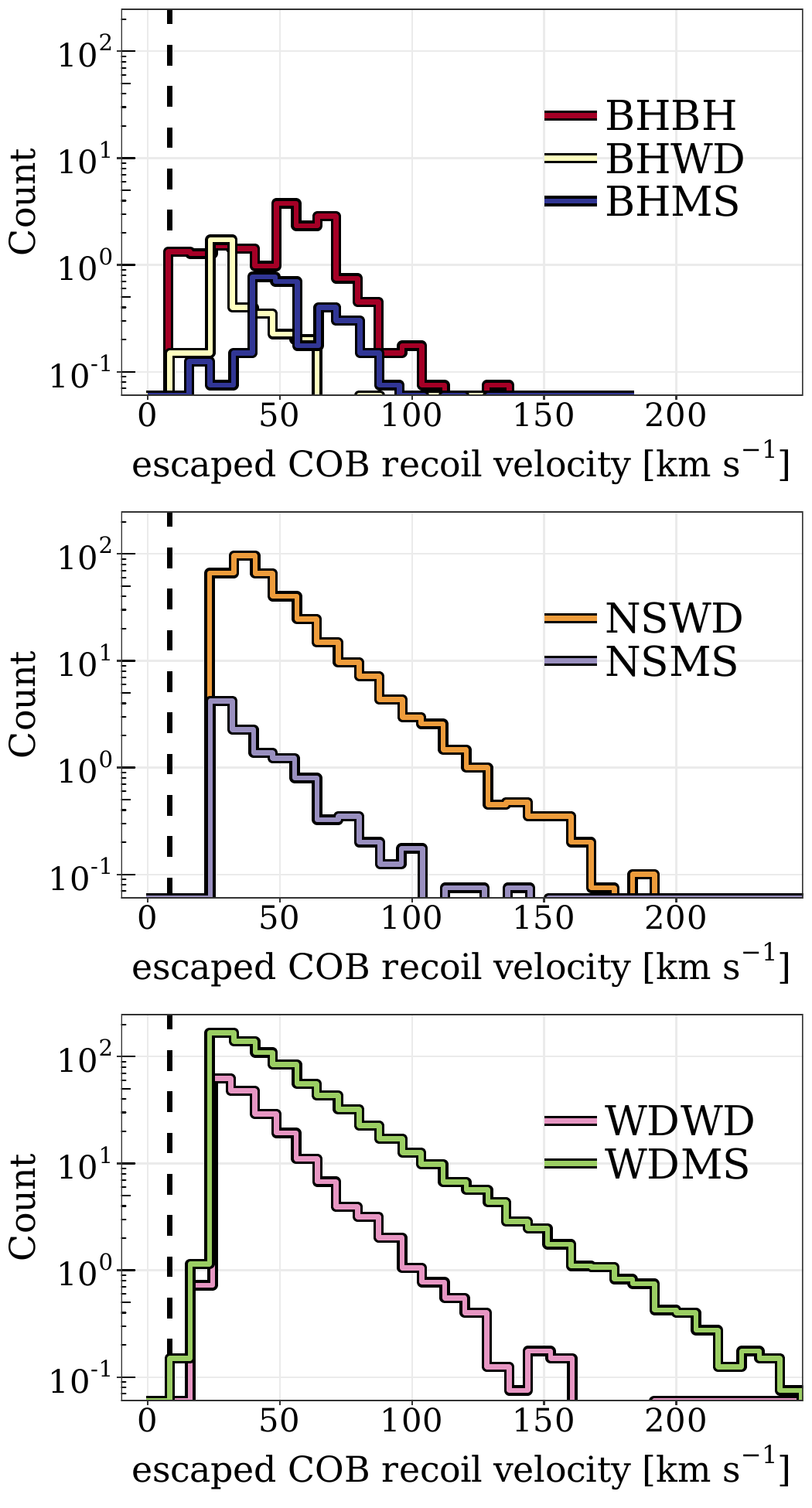}
    \caption{Recoil velocity distributions for all COBs which escape from their host GC following an S+COB encounter, sorted by COB type. Shown counts are averaged over fifty runs. The vertical line shows the minimum GC escape velocity. The distributions for BHNS and NSNS binaries are not shown because none have been ejected by any GC in the last 500 Myr. }
    \label{fig:vbin}
\end{figure}

\begin{table*}
    \begin{tabular}{|c|c|c|c|c|}
    
    \hline
    COB type & $N_{\rm COB, esc}$ & $v_{\rm rec, esc} \; \mathrm{[km \ s^{-1}}]$ & $\langle N_{\rm COB, esc}/ N_{\rm esc}\rangle $ & $\langle N_{\rm COB, esc} / N_{\rm enc, COB} \rangle$ \\ \hline 
    
    BHBH & $14_{-4}^{+5}$ & $50_{-29}^{+14}$ & $1.3_{-0.4}^{+0.4}  \times 10^{-2}$ & $6.8_{-2.0}^{+1.5}\times10^{-3}$ \\[0.1cm] 
    
    BHNS & 0 & - & - & - \\[0.1cm] 
    
    BHWD & $1_{-1}^{+4}$ & $25_{-3}^{+17}$ & $1.0_{-1.0}^{+4.0} \times 10^{-3}$ & $1.6_{-1.6}^{+5.8}\times10^{-3}$\\[0.1cm]
    
    BHMS & $2_{-2}^{+3}$ & $46_{-5}^{+24}$ & $1.9_{-1.9}^{+2.7} \times 10^{-3}$ & $1.3_{-1.3}^{+2.8}\times10^{-3}$\\[0.1cm]
    
    NSNS & 0 & - & - & - \\[0.1cm]
    
    NSWD & $273_{-59}^{+74}$ & $37_{-10}^{+20}$ & $2.6_{-0.5}^{+0.4}\times 10^{-1}$ & $2.2_{-0.1}^{+0.1}\times10^{-1}$\\[0.1cm]
    
    NSMS & $9_{-3}^{+4}$ & $32_{-9}^{+25}$ & $8.5_{-2.9}^{+5.2} \times 10^{-3}$  & $9.1_{-2.9}^{+7.0}\times10^{-2}$\\[0.1cm]
     
    WDWD & $140_{-15}^{+53}$ & $33_{-9}^{+19}$ & $1.4_{-0.1}^{+0.4}\times 10^{-1}$ & $2.2_{-0.1}^{+0.2}\times10^{-1}$\\[0.1cm]
    
    WDMS & $588_{-59}^{+35}$ & $40_{-14}^{+30}$ & $5.7_{-0.8}^{+0.5} \times 10^{-1}$ & $2.0_{-0.2}^{+0.1}\times10^{-1}$\\[0.1cm] 
    
    \hline
    \end{tabular}
    \caption{From left to right, the number of COB escapers in the last 500 Myr for each COB type, the typical recoil velocity of escaping COBs, the fraction of all escaped COBs that are each type, and the fraction of S+COB encounters of each COB type which result in an escaped COB.}
    \label{tab:binaries}
\end{table*}

Until now, we have explored the production of \textit{single} HVSs from 2+1 interactions in GC cores. During these encounters the COB receives a recoil velocity kick whose magnitude depends on the strength and specific geometry of the encounter. These kicks can impact the cluster COB populations; for instance, it is common for COBs containing a black hole to be ejected from clusters undergoing core collapse \citep{Kremer2019, Ye2019}. However, \texttt{CMC} simulations from \cite{Weatherford2023} find that three-body encounters are responsible for producing only $\sim$ 5\% of binary escapers (compared to two-body relaxation producing $\sim$80\% of binary escapers).

Yet, even if a binary is not fully ejected from a cluster, three-body interactions can play a large role in the spatial distribution of cluster binaries. Observations have discovered COBs that are significantly offset from their hosts -- For example, the core-collapsed GC NGC 6752 hosts a millisecond pulsar-WD binary located 3.3 half-light radii beyond the cluster centre \citep{DAmico2002}. While spatial offsets of this magnitude are typically explained via interactions with massive black holes, \cite{Leigh2024} recently showed that three-body interactions can be among the dominant mechanisms causing COBs like the one associated with NGC 6752 to migrate to the outskirts of clusters. Moreover, in a newly discovered observational sample of WDMS candidates in open clusters, \cite{Grondin2024WDMS} found that many of the WDMS candidates were located at distances far from their host clusters. While this study focused on a natal kick hypothesis to explain these spatial offsets, dynamical ejection mechanisms could also be present (albeit to a lesser degree in open clusters), causing the binary migration.

\texttt{Corespray} includes functionality to track the recoil velocity of the COBs, and therefore which COBs are kicked strongly enough to escape their host cluster. In Fig.~\ref{fig:vbin} we show the distributions of binary recoil velocities among COBs which have escaped their host cluster in the last 500 Myr. $18_{-4}^{+8}$ COBs containing a BH have escaped from GCs in the last 500 Myr, mostly with modest ($v \lesssim 50 \; \mathrm{km \ s^{-1}}$) recoil velocities. On the other hand, many more COBs containing a WD have escaped, which is not particularly surprising considering WDs are the most dominant CO in evolved GCs. $273_{-59}^{+74}$ NSWD binaries, $140_{-15}^{+53}$ WD-WD binaries, and $588_{-59}^{+35}$ WDMS binaries have escaped in the last 500 Myr over a large range of velocities. Of particular interest are the $31_{-5}^{+6}$ WDMS binaries with recoil velocities of $\geq 100 \; \mathrm{km \ s^{-1}}$. Our modelling even suggests the ejection of WD-MS binaries with recoil velocities above $150 \; \mathrm{km \ s^{-1}}$, though $<$1 such escapers are expected on average.  In Table \ref{tab:binaries} we offer some further statistics, including the fraction of all escaping COBs which are of each type and the fraction over \textit{all} COBs of each type which escape. WDMS systems constitute $\sim$57\% of all escaped COBs, NSWD binaries ($\sim$26\%) and WDWD binaries ($\sim$14\%) dominate the remainder of the escaped COB population. 
Finally, we include the fraction of all S+COB interactions of each COB type which result in the escape of the COB. While the probability of ejecting a binary including a BH from the cluster is low, $\sim10-20$\% of interactions with NSWD, WDWD or WDMS binaries result in the ejection of the binary.

\subsection{Intermediate mass black holes in Milky Way GCs}

The black holes in the COBs studied in this work are no more than a few tens of $\mathrm{M_\odot}$ in mass. The dense cores of GCs, however, are also promising locations to find \textit{intermediate mass black holes} (IMBHs) with masses from 100 $\mathrm{M_\odot}$ to $10^{5} \; \mathrm{M_\odot}$ \citep{Fujii2024}. The history of theoretical modeling and observational claims (and refutations)  of IMBHs in globular clusters is long \citep[see reviews in][]{Greene2020, Askar2023}. If an IMBH exists in a GC, Hills mechanism-like encounters between it and binary stars can eject stars at large velocities \citep{Fragione2019}. In practice, it would be difficult to distinguish stars ejected via S + COB encounters from those ejected by binary + IMBH encounters, and invoking the presence of an IMBH would be a tempting explanation for any fast star candidate with a trajectory which implies an origin in a Galactic GC \citep[see][]{Huang2025}.

To consider whether IMBHs might affect our predictions, we have searched the literature for evidence in support of or against IMBHs residing in the GCs listed in Table \ref{tab:counts}. We could not find published constraints on the existence of an IMBH in NGC 7099, NGC 6287 or NGC 6541. Combining FLAMES/VLT integral field spectroscopy with HST photometry, \citet{Lutzgendorf2013} derive upper limits on the mass of possible IMBHs in NGC 1851 and NGC 5824 of $2\times10^3 \; \mathrm{M_\odot}$ and $6\times10^3 \; \mathrm{M_\odot}$, respectively. While dynamical modeling studies of NGC 7078's centre have suggested the presence of an IMBH \citep{Newell1976, Gerssen2002}, follow-up studies have not found strong evidence in favour of an IMBH \citep{Murphy2011, Kirsten2012, denBrok2014}. NGC 6388 is the only cluster among these seven which remains suspected of harbouring an IMBH. This is still debated, however, as astrometric and photometric observations of \citet{Lanzoni2007} and spectroscopic observations of \citet{Lutzgendof2011, Lutzgendorf2015} suggest a central IMBH with a mass of $0.6-3\times10^4 \;\mathrm{M_\odot}$, while the spectoscopic observations of \citet{Lanzoni2013} and the radio observations of \citet{Cseh2010} and \citet{Tremou2018} seem to rule out an IMBH more massive than $\sim2000 \; \mathrm{M_\odot}$.

To conclude, in the seven Milky Way GCs which most contribute to the detectable or fast populations of S+COB stars, contamination by stars ejected following an IMBH interaction is not a concern (with the possible exception of NGC 6388).

\section{Conclusions} 
\label{sec:conclusions}

In this work we have investigated three-body interactions in Milky Way globular clusters between single stars and binaries that include at least one compact object. These \textit{star - compact object binary} (S+COB) interactions, though rare, are often energetic enough to eject single stars at velocities above the globular cluster escape velocity, and sometimes in excess of the escape velocity of the Galaxy itself ($\sim 500 \; \mathrm{km \ s^{-1}}$ at the Solar position). 
Our conclusions are as follows:

\begin{itemize}
    \item S+COB interactions in GC cores can eject single stars at several hundred $\mathrm{km \ s^{-1}}$, fast enough to escape GC cores. Our modelling suggests $6330_{-300}^{+340}$ stars have escaped from 100 clusters this way in the Galaxy in the past $500 \; \mathrm{Myr}$ (Figs. \ref{fig:vej}, \ref{fig:StairStep_cluster}, \ref{fig:StairStep_all}).
    \item $290_{-23}^{+28}$ of these stars are \textit{presently detectable} in the third data release from the \textit{Gaia} space mission with full five-parameter astrometry. However, only a maximum of two stars are predicted to be present in the \textit{Gaia} DR3 radial velocity catalogue. $1419_{-77}^{+116}$ stars are bright enough to be detectable in the near future in LSST (Fig.~\ref{fig:Skyall}).
    \item $839_{-67}^{+70}$ stars ejected via S+COB interactions in the last 500 Myr have present-day Galactocentric total velocities above $500 \; \mathrm{km \ s^{-1}}$. 22\% of these are ejected by a single GC, the dense cluster NGC 7078. However, only $1_{-1}^{+2}$ should be in \textit{Gaia} DR3 and $13_{-4}^{+3}$ will be detectable by LSST (Figs. \ref{fig:summaryfast}, \ref{fig:clusternumsfast}).
    
    \item One dense cluster in particular, NGC 7099, accounts for half of all stars detectable by \textit{Gaia} or LSST. Even so, identifying its ejected stars by \textit{Gaia} data alone is a challenging prospect (Fig. \ref{fig:70786397}). 
\end{itemize}

These results show that S+COB interactions in Milky Way GCs may be a more generous source of high-velocity and extratidal stars in the Galaxy than previously appreciated. The ejection of one hypervelocity star per $\sim$4.5 Myr from NGC 7099 is a rate at least two orders of magnitude smaller than the upper limit on the HVS ejection rate from the Galactic Centre via the Hills mechanism \citep[see][]{Marchetti2022, Verberne2024}. NGC 7099 might therefore be a minor but non-negligible contributor to the Galactic population of unbound stars. In other galaxies with more GCs and/or GCs more hospitable to S+COB interactions, this rate could be similar in magnitude to the Hills mechanism ejection rate. Of particular interest is M31, which has a population of hundreds of confirmed and candidate GCs \citep[see][]{Battistini1987, Barmby2000, Huxor2014, Wang2023}. HVSs ejected from M31 via the Hills mechanism can traverse the span between galaxies and pass into and through the Milky Way \citep{Sherwin2008, Gulzow2024} -- whether S+COB ejections in M31 also contribute to this population of Galactic intruders is an interesting question.

In addition to \textit{Gaia} DR4 and LSST, observational prospects will improve in the future with upcoming planned or proposed facilities and telescopes such as \textit{Roman} \citep{Spergel2015}, \textit{4MOST} \citep{deJong2019}, \textit{MOONS} \citep{Cirasuolo2020}, \textit{GaiaNIR} \citep{Hobbs2016} and \textit{THEIA} \citep{Theia2017} will offer further access to the faintest Milky Way stars. Searches for extratidal GC stars in the future will prove instrumental in studying globular cluster cores and the demographics of interesting stellar and non-stellar objects which lurk within them. 

\section*{Acknowledgements}
The authors are grateful to the organizers of the University of Toronto Local Group Hack Day, where this project originated. The authors thank Claude Cournoyer-Cloutier, Alison Sills, Newlin Weatherford, Holger Baumgardt, Ting Li, Vincent Hénault-Brunet, Abigail Battson, Maria Drout, and Ren\'{e}e Hlo\v{z}ek for helpful discussions and feedback. FAE acknowledges support from the University of Toronto Arts \& Science Postdoctoral Fellowship and Dunlap Postdoctoral Fellowship programs. SMG acknowledges the support of the Natural Sciences and Engineering Research Council of Canada (NSERC) and is partially funded through a NSERC Postgraduate Scholarship – Doctoral. SMG also recognizes funding from a Walter C. Sumner Memorial Fellowship. C.S.Y. acknowledges support from NSERC DIS-2022-568580. JB acknowledges financial support from NSERC (funding reference number RGPIN-2020-04712). A.L. acknowledges support from NSERC and is partially funded through a NSERC Canada Graduate Scholarship—Doctoral. A.L. is also supported by the Data Sciences Institute at the University of Toronto through grant number DSI- DSFY3R1P02. The Dunlap Institute is funded through an endowment established by the David Dunlap family and the University of Toronto.

\section*{Data Availability}
The simulations underpinning this work can be provided upon reasonable request to the corresponding author. The \texttt{Python} packages \texttt{speedystar} \citep{Contigiani2019, Evans2022b} and \texttt{corespray} \citep{Grondin2023} used to generate mock populations are publicly available.

\bibliographystyle{mnras}
\bibliography{HRS}

\label{lastpage}
\end{document}